\documentclass[fleqn,usenatbib]{mnras}

\usepackage{color}
\usepackage[T1]{fontenc}
\usepackage{ae,aecompl}
\usepackage{xcolor}
\usepackage{ulem}
\usepackage{graphicx}	% Including figure files
\usepackage{amsmath}	% Advanced maths commands
\usepackage{cases}
\usepackage{adjustbox}

\usepackage{graphicx}	% Including figure files
\usepackage{amssymb}	% Extra maths symbols
\usepackage{newtxtext,newtxmath}
\usepackage{cases}
\usepackage{array,multirow}
\usepackage{upgreek}
\usepackage{textcomp}
\usepackage{rotating}
\usepackage{fontawesome}
\newcommand{\Msol}{\;M_{\odot}}

\newcommand{\Rsol}{\;R_{\odot}}
\newcommand{\gram}{\;\mathrm{g}}
\newcommand{\cm}{\;\mathrm{cm}}

\newcommand{\km}{\;\mathrm{km}}
\newcommand{\AU}{\;\mathrm{AU}}
\newcommand{\pc}{\;\mathrm{pc}}
\newcommand{\yr}{\;\mathrm{yr}}
\newcommand{\mstar}{\;M_{\star}}
\newcommand{\Mbh}{M_{\bullet}}

\newcommand{\mesa}{{\small MESA}}

\usepackage{tikz}
\newcommand*\emptycirc[1][1ex]{\tikz\draw (0,0) circle (#1);} 
\newcommand*\halfcirc[1][1ex]{%
  \begin{tikzpicture}
  \draw[fill] (0,0)-- (90:#1) arc (90:270:#1) -- cycle ;
  \draw (0,0) circle (#1);
  \end{tikzpicture}}
\newcommand*\fullcirc[1][1ex]{\tikz\fill (0,0) circle (#1);} 

\defcitealias{Ryu+2022}{Paper 1}

\title[Transients in 3-body Encounters]{Close Encounters of Tight Binary Stars with Stellar-mass Black Holes}

\author[T. Ryu et al.]{
Taeho Ryu$^{1,2}$,\thanks{E-mail: tryu@mpa-garching.mpg.de}
Rosalba Perna$^{3,4}$,
Ruediger Pakmor$^{1}$,
Jing-Ze Ma$^{1}$,
Rob Farmer$^{1}$,
Selma E. de Mink$^{1,5}$
\\
$^{1}$ Max Planck Institute for Astrophysics, Karl-Schwarzschild-Strasse 1, 85748 Garching, Germany\\
$^{2}$ Physics and Astronomy Department, Johns Hopkins University, Baltimore, MD 21218, USA\\
$^{3}$ Department of Physics and Astronomy, Stony Brook
  University, Stony Brook, NY 11794-3800, USA\\
$^{4}$ Center for Computational Astrophysics, Flatiron Institute, New York, NY 10010, USA\\
$^{5}$ Anton Pannekoek Institute for Astronomy, University of Amsterdam, Science Park 904, 1098XH Amsterdam, The Netherlands
}

\date{Accepted XXX. Received YYY; in original form ZZZ}

\pubyear{2022}

\begin{document}
\label{firstpage}
\pagerange{\pageref{firstpage}--\pageref{lastpage}}
\maketitle

\begin{abstract}
Strong dynamical interactions among stars and compact objects are expected in a variety of astrophysical settings, such as star clusters and the disks of active galactic nuclei. Via a suite of 3D hydrodynamics simulations using the moving-mesh code AREPO, we investigate the formation of transient phenomena and their properties in close encounters between an $2M_{\odot}$ or $20M_{\odot}$ equal-mass circular binary star and single $20M_{\odot}$ black hole (BH). Stars can be disrupted by the BH during dynamical interactions, naturally producing electromagnetic transient phenomena. Encounters with impact parameters smaller than the semimajor axis of the initial binary frequently lead to a variety of transients whose electromagnetic signatures are qualitatively diﬀerent from those of ordinary disruption events involving just two bodies. These include the simultaneous or successive disruptions of both stars and one full disruption of one star accompanied by successive partial disruptions of the other star. On the other hand, when the impact parameter is larger than the semimajor axis of the initial binary, the binary is either simply tidally perturbed or dissociated into bound and unbound single stars (“micro-Hills” mechanism). The dissociation of $20M_{\odot}$ binaries can produce a runaway star and an active BH moving away from one another. Also, the binary dissociation can either produce an interacting binary with the BH, or a non-interacting, hard binary; both could be candidates of BH high- and low-mass X-ray binaries.  
Hence our simulations especially confirm that strong encounters can lead to the formation of the (generally difficult to form) BH low-mass X-ray binaries.
\end{abstract}


%%%%%%%%%%%%%%%%%%%%%%%%%%%%%%%

\section{Introduction}

Dynamical interactions between stars and compact objects in dense environments, such as star clusters, play a very important role in a variety of astrophysical settings. Many dynamical interactions redistribute energy between star cluster members which drives them towards equilibrium. Furthermore, binaries, an energy source in clusters, can be formed or destructed via dynamical interactions \citep{Hut1992}, which determines the cluster's thermodynamic state.

Occasionally, dynamical encounters between astrophysical objects lead them to interact at a close distance. In particular, close encounters involving stars and stellar-mass black holes (BHs) can often create transient phenomena, such as tidal disruption events (TDEs) \citep{Hills1988, Perets2016,Lopez2019,Kremer2019,Wang2021TDE,Kremer2021,Kremer2022}. In those the BH can fully or partially disrupt a star when the closest point of approach is smaller than the so-called tidal disruption radius $r_{\rm t}\sim (M_{\bullet}/M_{\star})^{1/3}R_{\star}$. Here, $M_{\bullet}$ is the black hole mass, $M_{\star}$ the stellar mass and $R_{\star}$ the stellar radius. In this process, a bright electromagnetic flare can be generated. In particular, encounters involving multiplets (e.g., binary), which frequently occur near the center of clusters, can create wider varieties of transients than those between two single objects because of the chaotic nature of interactions \citep{Lopez+2019,Ryu+2022}. 

Another X-ray source that can form during dynamical interactions between stars and black holes is an X-ray binary \citep[e.g.,][]{Kremer+2018,Kremer+2018c}, which can show a transient behavior via, e.g., a disk instability \citep[e.g.,][]{KingBurderi1996}. A number of BH X-ray binaries has been detected in our Galaxy \citep{Corral-Santana+2016} via radio \citep[e.g.,][]{Chomiuk+2013} or X-ray measurements \citep[e.g.,] []{Miller-Jones+2015,Shishkovsky+2018}. Among these, the detection of two dozens of \textit{BH low-mass X-ray binaries} (BH-LMXBs) \citep{Casares+2014} has been puzzling astronomers, and the formation channel of such systems has been debated. In the ``standard'' scenario, the orbit of a binary star system shrinks via a common envelope phase before one star collapses to a compact object \citep[e.g.,][]{deKool1987}. However, this scenario has been challenged by \citet{Podsiadlowski+2003}, who suggested that binaries with a very large mass ratio would eventually merge during the common envelope phase. Also dynamically-formed binaries via 3-body encounters tend to have comparable-mass companions as a result of a member exchange. Therefore, alternative scenarios have been examined, such as dynamical formation via a number of weak encounters \citep{MichaelyPerets2016} and formation in hierarchical triples \citep{Naoz+2016}. In addition to these formation channels, a strong three-body encounter between a star and black hole can form such systems. The last possibility adds to the importance of investigating the hydrodynamics of three-body encounters between stars and black holes.

Theoretical investigations of transient phenomena created in multi-body encounters are especially timely in light of the dramatically increase of the number of detectable transients with both ongoing surveys, such as eROSITA\footnote{https://erosita.mpe.mpg.de} and the Zwicky Transient Facility (ZTF)\footnote{https://www.ztf.caltech.edu}, but especially with
the upcoming Vera Rubin Observatory (VRO)\footnote{https://www.lsst.org}. Despite the impending increase of the number of transient candidates, there have been only a few attempts to perform hydrodynamics calculations for dynamical encounters involving binaries \citep[e.g.][]{McMillan+1991,GoodmanHernquist1991,Lopez+2019}. In \citet{Ryu+2022} (\citetalias{Ryu+2022} hereafter), we  explored the outcomes of close three-body encounters between a single main-sequence star and a merging (i.e. with lifetime shorter than the Hubble time) binary black hole using 3D smoothed particle hydrodynamics simulations. In particular, we focused on quantifying the impact of the close encounter, frequently resulting in the disruption of the star, on the binary orbit and compared that with what the point-particle approximation predicts. We showed that a single disruption event can change the binary orbital parameters up to 20\% of their pre-encounter values, or equivalently change the gravitational-wave-driven merger timescale up to order unity. This impact is often different from that of a pure scattering. We further showed that the accretion rates of both BHs that have undergone a disruption event are typically super-Eddington with modulations on a time scale of the binary orbital period.

We continue our investigation on three-body encounters between stars and stellar-mass BHs using hydrodynamics simulations. In particular, in this paper we focus on possible transient formation in close three-body encounters between a binary star and a single BH, using the moving-mesh hydrodynamics code {\small AREPO} \citep{Arepo}. This type of encounters can naturally happen in dense stellar environments where binaries exist, such as  open clusters, which have a binary fraction $\simeq 0.2-0.7$ \citep{Moe2017,Sollima+2010}, 
nuclear star clusters \citep{Fragione2021}, globular clusters \citep{Perets2016,Kremer2019}, young star clusters \citep{Kremer2021}, the disks of Active Galactic Nuclei \citep{Yang2022}. 
In addition to varieties of transient phenomena qualitatively different from ordinary TDEs, our numerical investigation shows that the outcome of close encounters between BHs and binary stars can lead to a variety of astrophysical outcomes, including the formation of both BH high- and low-mass X-ray binary and runaway stars with active single BHs, among others.

Our paper is organized as follows: Our numerical methods, based on the use of the moving mesh code {\small AREPO} are detailed in \S~\ref{sec:method}. Our results are presented in \S~\ref{sec:result}. In \S~\ref{sec:discussion}, we discuss their astrophysical implications (\S~\ref{sec:implication}) and estimate the rate (\S~\ref{subsec:rate}). We summarize and conclude in \S~\ref{sec:summary}.

\section{Simulation details}\label{sec:method}

In this work we investigate the outcome properties of nearly parabolic encounters between a binary star and a single stellar-mass black hole. In particular, we focus on the properties of the surviving remnants and on the accretion and spin state of the BH following the encounter.

\subsection{Numerical Methods}

To achieve our scientific goal, we perform a suite of 3D hydrodynamic simulations of the close encounters using the moving mesh code {\small AREPO} \citep{Arepo,ArepoHydro,Arepo2}. {\small AREPO} is a massively parallel gravity and magnetohydrodynamic code, which has been used for many astrophysical problems \citep[e.g.,][]{Illustris}. It adopts a second order finite-volume scheme to discretize the hydrodynamic equations on a moving Voronoi mesh, and a tree-particle-mesh method for gravitational interactions. This approach of constructing grids represents a compromise between the two widely used hydrodynamics schemes, that is the Eulerian finite-volume method and the Lagrangian smoothed particle method. As a result, this new approach inherits advantages of both schemes, such as improved shock capturing without introducing an artificial viscosity, and adaptive adjustment of spatial resolution. 
The gas self-gravity is computed in {\small AREPO} using a tree solver and is coupled to the hydrodynamics via a Leapfrog time integration scheme. 

We use the {\small HELMHOLTZ} equation of state \citep{HelmholtzEOS} which includes the radiation pressure, assuming local thermodynamic equilibrium. We include $8$ isotopes ($\mathrm{n}$, $\mathrm{p}$, $^{4}\mathrm{He}$, $^{12}\mathrm{C}$, $^{14}\mathrm{N}$, $^{16}\mathrm{O}$, $^{20}\mathrm{Ne}$, $^{24}\mathrm{Mg}$) with a nuclear reaction network \citep{Pakmor+2012}.

\begin{figure}
	\centering
	\includegraphics[width=8.6cm]{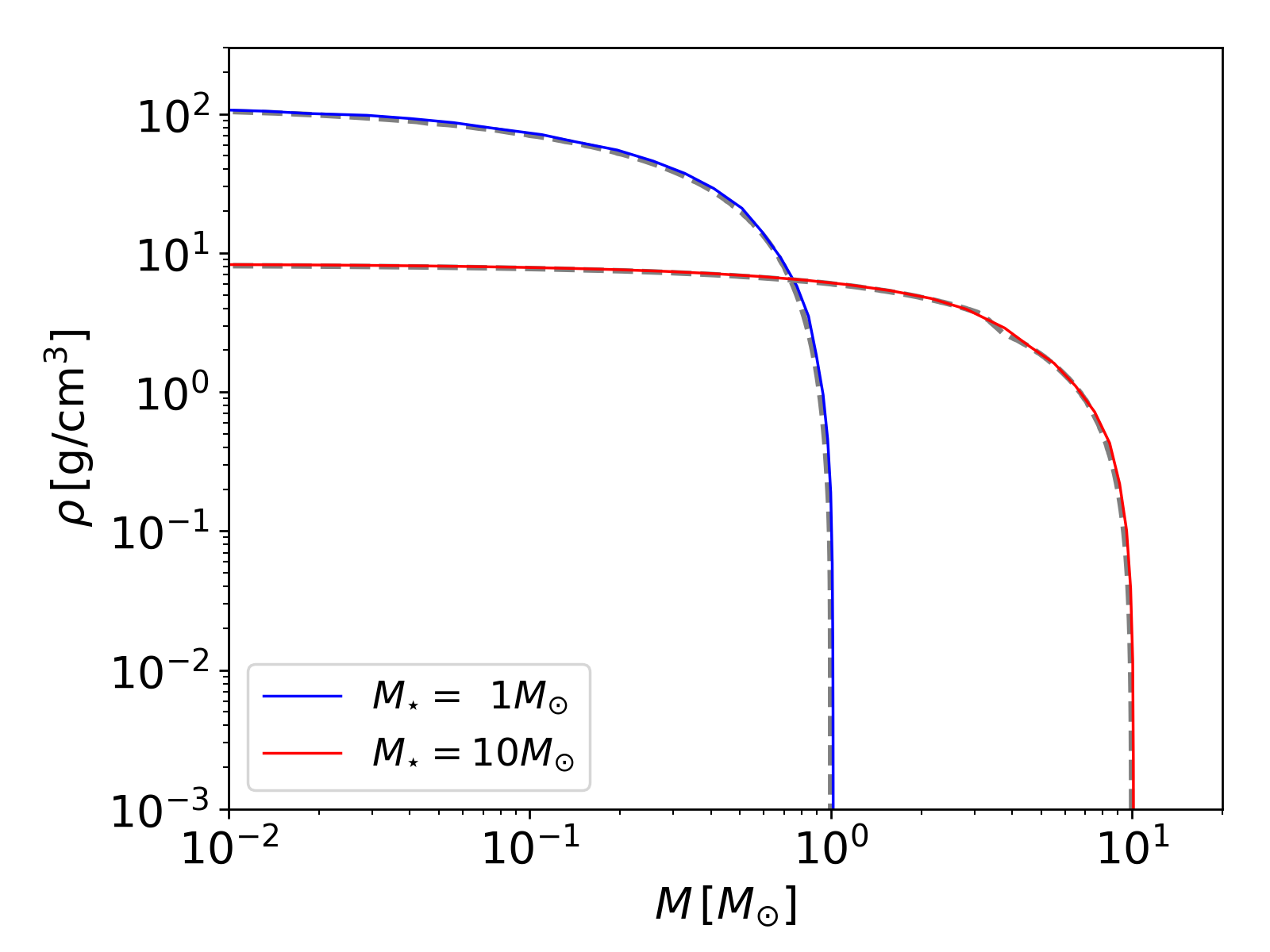}
\caption{The radial density profile of the main-sequence stars with $M_{\star}=1\Msol$ (blue) and $10\Msol$ (red) relaxed for five stellar dynamical times, as a function of mass. The dashed lines indicate the profiles for the {\sc MESA} models, which are just sitting below the solid lines.}
	\label{fig:density}
\end{figure}

\subsection{Binary stars}

The initial state of the stars was taken from somewhat evolved main-sequence (MS) stars (with the core H mass fraction of 0.5) computed using the stellar evolution code {\sc MESA} (version r22.05.1) \citep{Paxton+2011,paxton:13,paxton:15,MESArelaxation,paxton:19,jermyn22}. We map the 1D {\sc MESA} model into a 3D {\small AREPO} grid with $N\simeq 3\times10^{5}$ cells, using the profiles of density, pressure and chemical composition \citep{Ohlmann+2017}. The single star is first relaxed for five stellar dynamical times $t_{\rm dyn}=\sqrt{R_{\star}^{3}/GM_{\star}}$ where $R_{\star}$ and $M_{\star}$ are the radius and mass of the star, respectively. The density profiles of the relaxed stars considered in our simulations are depicted in Figure~\ref{fig:density}. 

We then relax binaries consisting of two relaxed single stars for $0.5 P$ where $P=2\pi \sqrt{a^{3}/G(M_{1}+M_{2})}$ is the period of the binary with the primary mass $M_{1}$, secondary mass $M_{2}$ and semimajor axis $a$. We parameterize the semimajor axis $a$ using an approximate analytic estimate of the Roche lobe radius \citep{Eggleton1983}, 
\begin{align}
   \frac{r_{\rm RL}}{a}= \frac{0.49 q^{2/3}}{0.6q^{2/3}+\ln(1+q^{1/3})},
\end{align}
where $r_{\rm RL}$ is the Roche lobe radius and $a$ is the orbital orbital separation.
For $q = 1$ and $r_{\rm RL} = R_{\star}$, we define $a_{\rm RL} \equiv a(R_{\rm RL}=R_{\star})$ as the separation at which both stars fill their Roche lobes. It follows that $ a_{\rm RL} \simeq 2.64R_{\star}$ and  $P/t_{\rm dyn}\simeq 19 (a/a_{\rm RL})^{3/2}$. 

We performed this binary relaxation process for every binary with different orbital parameters (6 different binaries in total). The semimajor axis and the eccentricity of the relaxed binaries differ by less than 1\% of their initial values. 

Note that we performed convergence tests with different number of cells ($N=2.5\times10^{5}$, $5\times10^{5}$, $10^{6}$ and $2\times10^{6}$) for a few cases in which we compared the outcomes and a few key quantities such as the semimajor axis and eccentricity of the final binary and black hole ejection speed. We found that all the simulations showed converged results.

\subsection{Black holes}

We model the BH using an initially non-rotating sink particle which interacts via gravity with gas and grows in mass via accretion of gas. The gravitational softening length of the BH is set to be the same as the minimum softening length of the cells of the stars. At every time step, we assume that the accretion rate is determined by the average inward radial mass flux towards the BH. More precisely, accretion occurs in four steps:
\begin{enumerate}
    \item \textit{Cell identification for mass flux calculation}: the code identifies cells around the BH within $1.5\times 10^{4}r_{\rm g}$ where $r_{\rm g} = G M_{\bullet}/c^{2}$ is the gravitational radius of the BH. We choose a somewhat large search radius to ensure the sampling of a large enough number of cells. However, as explained in ii), we compensate for the large search radius by adopting an inverse-distance weight function. 
    
    \item \textit{Accretion calculation}: to account for the fact that gas closer to the BH is expected to contribute more to the  accretion rate, we estimate  an average radial mass flux of the identified cells for given position and velocity of the BH as,
\begin{align}
    F^{r} = \sum_{i} m_{i} W(r_{i}, h_{i}) v_{i}^{r},
\end{align}
where $m_{i}$ is the mass of the selected neighboring cells, $W(r_{i}, h_{i})$ the (inverse-distance weighted) spline Kernel \citep{MonaghanLattanzio1985} (Equation 4 in \citealt{GADGET2}), $r_{i}$ the distance from the BH, $h_{i}$ the softening length of the BH, $v_{i}^{r}$ the velocity of the gas relative to the BH in the radial direction. The effective surface area $A$ is determined using the weighted average volume $V$ of the cells with volume $V_{i}$, $V = \sum_{i} V_{i} W(r_{i}, h_{i})/ \sum_{i} W(r_{i}, h_{i})$, giving $A=3^{2/3}(4\pi)^{1/3} V^{2/3}$ (assuming spherical geometry). Finally, the accretion rate is $\dot{M}=\mathrm{Max}(-F^{r}A,0) $. At every time step with step size $\Delta t$ where the BH's position and velocity are updated, the BH grows in mass by an amount of $\Delta M =(1-\eta)\dot{M}\Delta t$ where $\eta$ is the radiative efficiency. We assume $\eta = 0$ in our simulations.

    \item \textit{Mass subtraction and back reaction}: to conserve the mass, the same amount of mass is subtracted from a cell closest to the BH among those with a mass of at least $>5\Delta M$ and density $> 10^{-5}\gram\cm^{-3}$, and bound to the BH. Although the minimum mass is chosen mostly to ensure to avoid a very small mass of the cell after the mass subtraction, this particular choice should not affect our results because the masses of the subtracted cells are typically much above the minimum mass. The lower bound of the density is to ensure that the mass is subtracted from ``real'' gas, not from a vacuum cell. Then, in order  to properly take into account the back-reaction of accretion, we subtract the average momentum of the accreted mass,
\begin{align}
    \textbf{P} = \Delta M \frac{\sum_{i}  W(r_{i}, h_{i}) \textbf{v}_{i}}{\sum_{i} W(r_{i}, h_{i})},
\end{align}
where $\textbf{v}_{i}$ is the velocity vector of the $i_{\rm th}$ neighboring cell identified at the step i), from the ``mass-losing'' cell, and add the same amount to the BH. 

\item  \textit{Spin evolution}: accretion of gas can increase the BH spin. We evolve the BH spin due to mass accretion following \citetalias{Ryu+2022}, who adopted the formalism by \citet{Fanidakis+2011}. The only significant difference is how the direction of the accreted angular momentum is calculated: \citetalias{Ryu+2022}, which performed smoothed particle hydrodynamics simulations, tracked the angular momentum vector simply by cumulatively adding the momentum of the accreted particles, but in this work, we make use of the average angular momentum (measured in the BH frame) of the selected neighboring cells. 
\end{enumerate}

We should note that our simulations do not include the radiation feedback produced by accretion. This may contribute to create significant outflows and regulate the subsequent accretion, especially at super-Eddington accretion rates (e.g. \citealt{Skadowski2014}).
While our simulations reveal a radiation pressure gradient built up in the optically thick gas near the BH, which can push gas away from the BH, its impact is found to be small. 
 Given the purpose of this work, namely, to identify all possible transient types during star-binary/BH three-body encounters, and properly classify the outcomes, we defer improvements of the treatment of accretion and its feedback on surrounding gas to future work specifically dedicated to studying the outcome observables.
 
\begin{figure*}
	\centering
	\includegraphics[width=8.6cm]{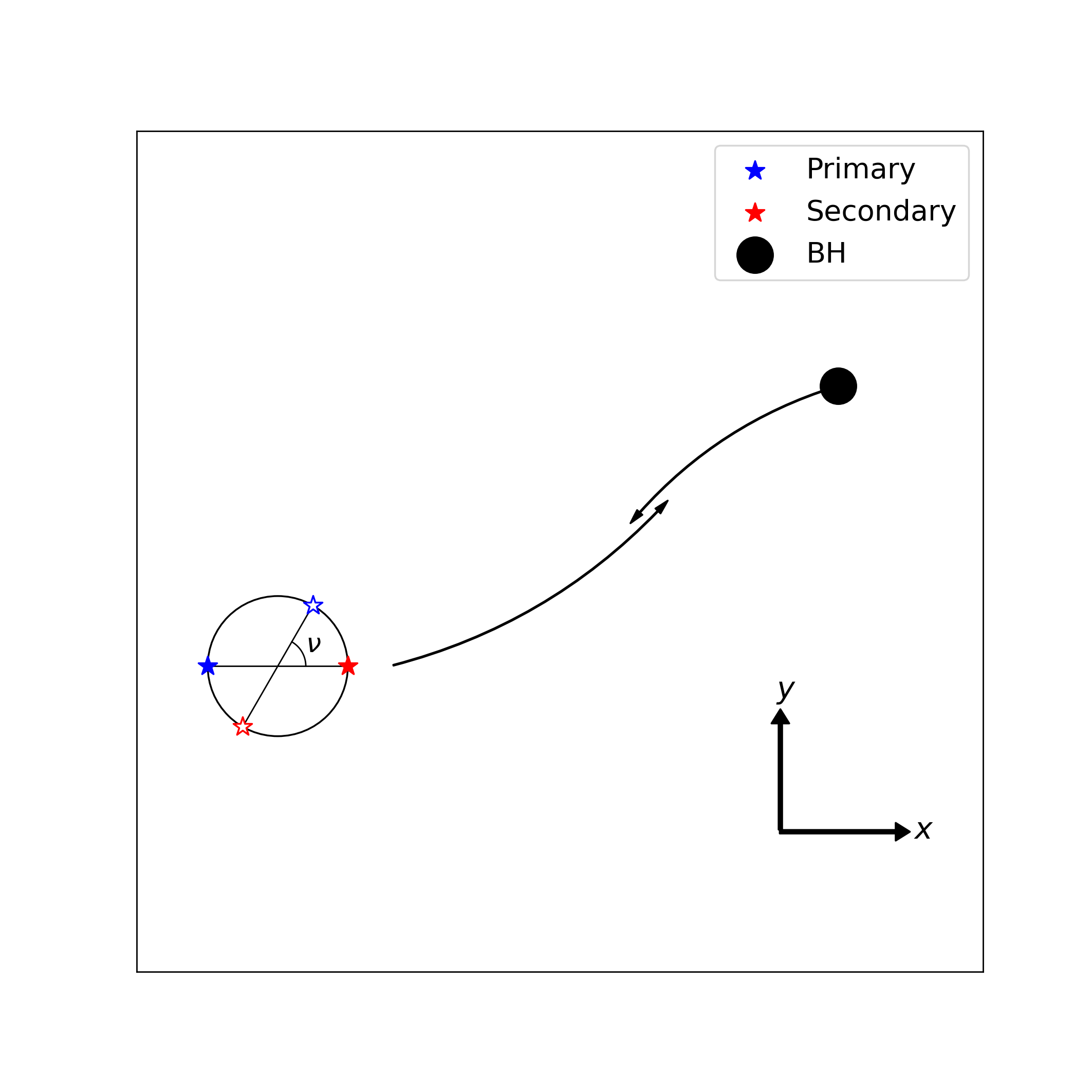}
		\includegraphics[width=8.6cm]{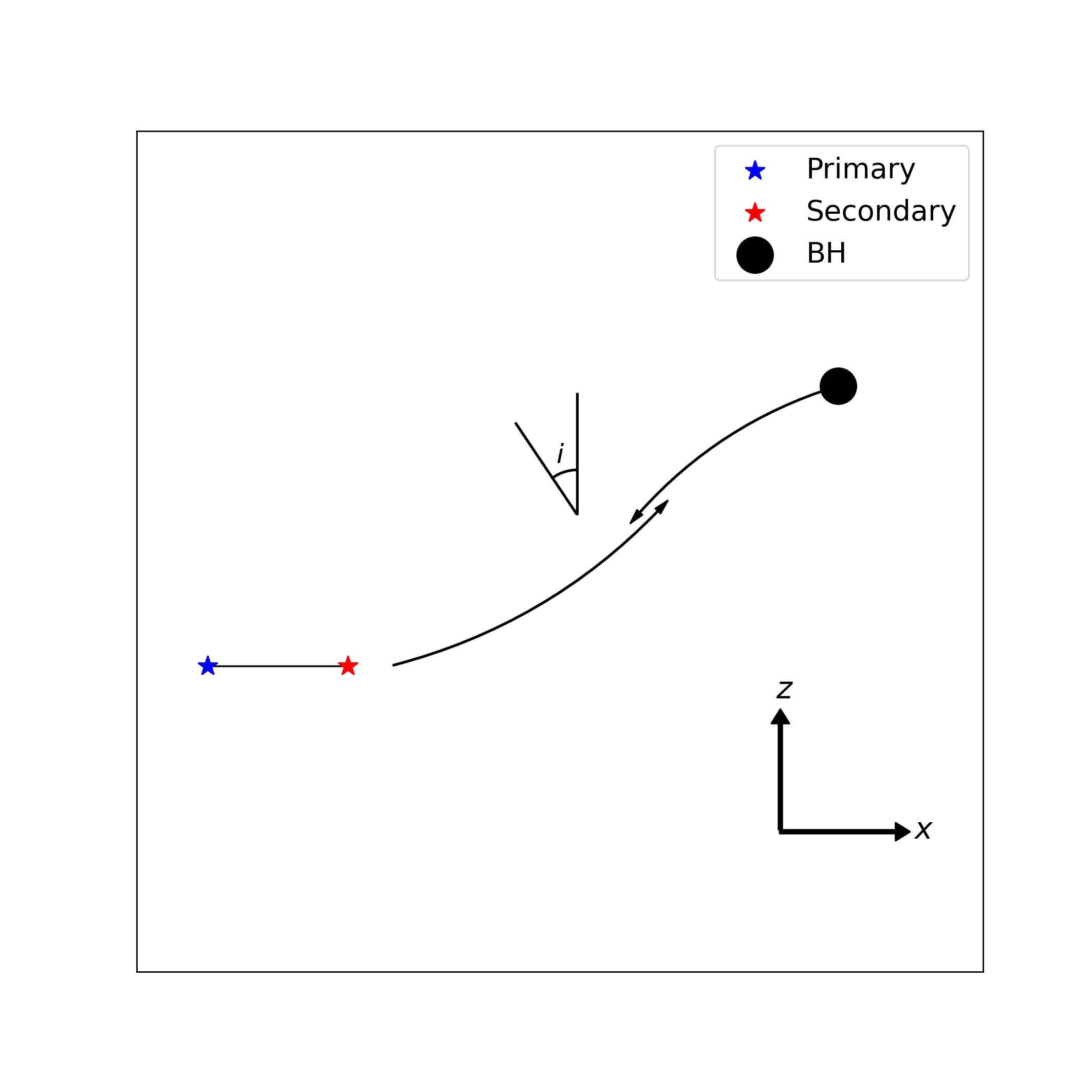}
\caption{Schematic diagrams for the initial configuration of the stellar binary (blue and red solid stars) and black hole (black circle) for a prograde case with the inclination angle $i<90^{\circ}$ and the phase angle $\nu=0^{\circ}$, projected onto the $x-y$ plane (\textit{left}) and the $x-z$ plane. The arrows indicate the instantaneous direction of motion. The open stars, on the same circle with the closed stars, indicate the case with $\nu >0^{\circ}$.}
	\label{fig:init}
\end{figure*}

\begin{table}
\begin{tabular}{  c c c c c c c c c c} 
\hline
Model name & $M_{b}$ & $a/a_{\rm RL}$ &  $i$  & $\upsilon~[^{\circ}]$  & $\beta$ & $t_{\rm p}$\\
\hline
Unit & $M_{\odot}$  & - & $^{\circ}$ & $^{\circ}$ & - & hours\\
\hline
$M2a6\beta2i30$ & 2  & 6 & 30 & 0 & 2 & 17\\
$M2a6\beta1i30$ & 2  & 6 & 30 & 0 & 1 & 6.1\\
$M2a6\beta1/2i30$ & 2  & 6 & 30 & 0 & 1/2 & 2.2\\
$M2a6\beta1/4i30$ & 2  & 6 & 30 & 0 & 1/4 & 0.76 \\
$M2a6\beta2i150$ & 2  & 6 & 150 & 0 & 2 & 17\\
$M2a6\beta1i150$ & 2  & 6 & 150 & 0 & 1 & 6.1\\
$M2a6\beta1/2i150$ & 2  & 6 & 150 & 0 & 1/2& 2.2 \\
$M2a6\beta1/4i150$ & 2  & 6 & 150 & 0 & 1/4 & 0.76 \\
\hline
$M2a2\beta1/2i30$ & 2  & 2  & 30 & 0 & 1/2  & 0.42 \\
$M2a2\beta1/2i150$ & 2  & 2 & 150 & 0 & 1/2  & 0.42 \\
$M2a9\beta1/2i30$ & 2  & 9  & 30 & 0 & 1/2  & 4.0\\
$M2a9\beta1/2i150$ & 2  & 9 & 150 & 0 & 1/2  & 4.0 \\
\hline
$M2a6\beta1/2i0$ & 2  & 6  & 0 & 0 & 1/2   & 2.2\\
$M2a6\beta1/2i60$ & 2  & 6  & 60 & 0 & 1/2  & 2.2 \\
$M2a6\beta1/2i120$ & 2  & 6  & 120 & 0 & 1/2  & 2.2 \\
$M2a6\beta1/2i180$ & 2  & 6 & 180 & 0 & 1/2   & 2.2\\
\hline
$M2a6\upsilon45$ & 2  & 6 & 30 & 45 & 1/2  & 2.2 \\
$M2a6\upsilon90$ & 2  & 6 & 30  & 90 & 1/2  & 2.2\\
$M2a6\upsilon135$ & 2  & 6 & 30 & 135 &  1/2  & 2.2 \\
\hline
\hline
$M20a6\beta2i30$ & 20  & 6 & 30 & 0 & 2 & 50\\
$M20a6\beta1i30$ & 20  & 6 & 30 & 0 & 1 & 18\\
$M20a6\beta1/2i30$ & 20  & 6 & 30 & 0 & 1/2 & 6.3 \\
$M20a6\beta1/4i30$ & 20 & 6 & 30 & 0 & 1/4 & 2.3\\
$M20a6\beta2i150$ & 20  & 6 & 150 & 0 & 2  & 50\\
$M20a6\beta1i150$ & 20  & 6 & 150 & 0 & 1& 18 \\
$M20a6\beta1/2i150$ & 20  & 6 & 150 & 0 & 1/2 & 6.3 \\
$M20a6\beta1/4i150$ & 20  & 6 & 150 & 0 & 1/4& 2.3 \\
\hline
$M20a2\beta1/2i30$ & 20 & 2  & 30 & 0 & 1/2  & 1.2  \\
$M20a2\beta1/2i150$ & 20  & 2 & 150 & 0 & 1/2  &1.2  \\
$M20a9\beta1/2i30$ & 20  & 9  & 30 & 0 & 1/2  & 12  \\
$M20a9\beta1/2i150$ & 20  & 9 & 150 & 0 & 1/2  & 12  \\
\hline
$M20a6\beta1/2i0$ & 20  & 6  & 0 & 0 & 1/2 & 6.4 \\
$M20a6\beta1/2i60$ & 20  & 6  & 60 & 0 & 1/2 & 6.4 \\
$M20a6\beta1/2i120$ & 20  & 6  & 120 & 0 & 1/2 & 6.4 \\
$M20a6\beta1/2i180$ & 20  & 6 & 180 & 0 & 1/2 & 6.4  \\
\hline
\end{tabular}
\caption{The initial model parameters. The model name (first column) contains the information of key initial parameters: for the model names of $M(1)a(2)\beta(3)i(4)$, (1) binary mass $M_{\rm b}$ 
 [$\Msol$], (2) the initial semimajor axis of the binary $a/a_{\rm RL}$, (3) the impact parameter $\beta$ and (4) the inclination angle $i$ in degrees, and for the model names of $M(1)a(2)\nu(3)$, (1) binary mass $M_{\rm b}$ [$\Msol$], (2) $a/a_{\rm RL}$, (3) the initial phase angle $\nu$ in degrees (see Figure~\ref{fig:init}).  Note that $a_{\rm RL} \simeq 2.64R_{\star}$. $t_{\rm p}$ is the dynamical time at $r=r_{\rm p}$, in units of hours, and $r_{\rm p}$ is the pericenter distance of the incoming orbit in hours relative to the BH $r_{\rm p}=\beta[M_{\bullet}/(M_{1}+M_{2})]^{1/3}(1+e)a/2$. }\label{tab:initialparameter}
\end{table}

\subsection{Initial parameters}\label{sub:initialparameter}

Throughout the paper, quantities with the subscript containing $b(\star)-\bullet$ refer to those related to the orbit between a binary (single star) and the BH. We consider a parabolic encounter with  eccentricity $1-e_{b-\bullet}=10^{-5}$. Parabolic orbits are a reasonable assumption for encounters in star clusters given that the typical eccentricity of two-body encounters in a cluster with velocity dispersion $\sigma$ is $|1-e|\simeq 10^{-4}(\sigma/15\km{\rm s}^{-1})^{2}(\Mbh/40\Msol)^{-2/3}(M_{\star}/1\Msol)^{-1/3}(R_{\star}/1\Rsol)$ for $\Mbh\gg \mstar$\footnote{ The situation is somewhat different in AGN disks, where during the early times, when prograde orbits are in the process of being damped, there is a much higher likelihood of encounters being hyperbolic \citep{Secunda2021}. However, as time goes on, gentler encounters are expected. }.
The distance between the binary's center of mass and the BH at the first closest approach $r_{\rm p, b-\bullet}$ is parameterized using the impact parameter $\beta$, i.e., $r_{\rm p,b-\bullet}=0.5\beta r_{\rm t}$. Here, $r_{\rm t}$ is defined as $[M_{\bullet}/(M_{1}+M_{2})]^{1/3} a (1+e)$ where $a$ and $e$ are the binary semimajor axis and eccentricity, respectively. 

For the mass of the BH we choose $20M_{\odot}$ \citep[c.f.,][]{SperaMapelli2017}.
To study the impact of binary mass on outcomes, we consider a low-mass case, where we choose $M_{\rm b} = M_{1}+ M_{2}=2\Msol$, ($M_{1,2}=1\Msol$) and a high mass case where we chose $M_{\rm b} = M_{1}+ M_{2}=20\Msol$, ($M_{1,2}=10\Msol$). We consider three semi-major axes: $a/a_{\rm RL}=2$, $6$ and $9$, and assume the orbit to be circular initially. Massive stars are commonly found in such close binary systems \citep{Sana+2012}. For lower-mass stars such tight systems are rare \citep{DucheneKraus2013}, but their longer life lifetimes and the fact that they are favored by the initial mass function may still make possible encounters of these low-mass binaries with a BH relevant to explore. The choice of an initially circular binary orbit may be justified by the fact that tidal circularization timescales are shorter for more compact binaries. In fact, \citet{Meibom+Mathieu2005} found that binaries in open clusters are circularized out to $P\simeq 8 - 15$ days, which is comparable to or longer than the binary period considered in this study. Nonetheless, three-body encounters involving eccentric binaries are also possible. Because the effective encounter cross section is greater for eccentric orbits (by a factor $\simeq 1+e$), the overall encounter rate would be higher, but the impact of other parameters (e.g., semimajor axis, impact parameters, see \S\ref{subsec:dependence}) on the final outcome would be dominant over that of different initial binary eccentricities.

The binary's angular momentum axis is always along the $z$ axis in our simulations. This configuration defines the mutual inclination angle, which is illustrated in Figure~\ref{fig:init} showing the initial configuration of the stellar binary and black hole. We examine the outcomes of encounters with various values of $i$ and $\beta$: $i=0$, $30^{\circ}$, $60^{\circ}$, $120^{\circ}$ and $180^{\circ}$, and $\beta = 1/4$, 1/2, 1 and 2. However, given the relatively high computational costs, we do not simulate encounters with every combination of $i$ and $\beta$. Instead, we simulate the encounters of the intermediate-size binaries ($a/a_{\rm RL}=6$) with different combinations of $\beta=1/4$, 1/2, 1 and 2, and $i=30^{\circ}$, $150^{\circ}$. For the smallest and largest binaries ($a/a_{\rm RL}=2$ and 9), we only consider $i=30^{\circ}$ and $150^{\circ}$ while $\beta=1/2$. In addition, we further examine the dependence of $i$ on the outcome properties by considering $i=0$, $60^{\circ}$, $120^{\circ}$ and $180^{\circ}$ (for $\beta=1/2$). Last, we also study the impact of the phase angle $\nu$ (see Figure~\ref{fig:init}) on the encounter outcomes. We define $\nu$ as the initial angle between the line connecting the two stars in the binary and the $x-$axis. To simulate encounters with different phase angles ($\nu = 45^{\circ}$, $90^{\circ}$ and $135^{\circ}$), we initially place the binary with a different phase angle while all other parameters remains fixed.

The initial separation between the binary and BH is $5r_{\rm t}$. 

We summarize the initial parameters considered in our simulations in Table~\ref{tab:initialparameter}.
Each of the models is integrated up to $> 25 (130) t_{\mathrm p}$ for the encounters of the $2 (20)\Msol$ binary which it takes to identify the final outcomes. Here, $t_{\rm p}= \sqrt{r_{\rm p}^3/G(M_{\bullet}+M_{\rm b})}$ is the dynamical time at $r=r_{\rm p}$. The values of $t_{\rm p}$ for each model is given in Table~\ref{tab:initialparameter}.

\begin{table*}
\resizebox{\textwidth}{!}{
\begin{tabular}{  c| c |c  c c c c c c |c c c c c c c| c} 
\hline
\multirow{2}{*}{Model name} & \multirow{2}{*}{Class} & \multicolumn{7}{c|}{Outcome 1} & \multicolumn{7}{c|}{Outcome 2} & BH\\
                           &                       & Type & $\Delta M/M$ & $a[R_{\odot}]$ & $e$ & $r_{\rm p}[R_{\odot}]$ & $P$[yr] & $v$ & Type & $\Delta M/M$ & $a[R_{\odot}]$ & $e$ & $r_{\rm p}[R_{\odot}]$ & $P$[yr] & $v$ [km/s] & $v$ [km/s]\\
\hline
$M2a6\beta2i30$ & Non-disruptive & \fullcirc $^{\star}$ & -0.21 & 60 &  0.83 &  9.8 &  0.03 &  - &  \fullcirc & -0.12 &  - &  - &  - &  - & 220 &  -\\%9.7\\
$M2a6\beta1i30$ & Disruptive &  \fullcirc $\longrightarrow$ \halfcirc$^{\star}$ & -0.26 & 85 &  0.95 &  3.9 &  0.05 &  - &  \fullcirc & -0.14 &  - &  - &  - &  - & 170. & -\\%14\\
$M2a6\beta1/2i30$ & Disruptive &  \halfcirc  $\longrightarrow$  \emptycirc & -25 & 78 &  0.98 &  1.3 &  0.05 &  - &  \fullcirc & -0.18 &  - &  - &  - &  - & 180 & -\\%23\\
$M2a6\beta1/4i30$ & Disruptive &  \emptycirc &  - &  - &  - &  - &  - &  - &  \fullcirc & -0.39 &  - &  - &  - &  - & 160 & 14\\
$M2a6\beta2i150$ & Non-disruptive &  - &  - & 16 &  0.69 &  4.8 &  0.01 &  - &  - &  - & 7400 &  1.0 & 24 & 140 &  - & 25\\
$M2a6\beta1i150$ & Non-disruptive &  \fullcirc$^{\star}$ & -0.16 & 300 &  0.93 & 22 &  0.36 &  - &  \fullcirc & -0.24 &  - &  - &  - &  - & 110 & -\\%23\\
$M2a6\beta1/2i150$ & Non-disruptive &  \fullcirc & -0.34 & 200 &  0.99 &  2.5 &  0.19 &  - &  \fullcirc & -0.22 &  - &  - &  - &  - & 120 & -\\%23\\
$M2a6\beta1/4i150$ & Disruptive &  \halfcirc & -3.1 &  - &  - &  - &  - & 180 &  \halfcirc  $\longrightarrow$  \emptycirc & -5.8 & 63 &  0.99 &  0.85 &  0.03 &  - & -\\%14\\
\hline
$M2a2\beta1/2i30$ & Disruptive &  \emptycirc & - &  - &  - &  - &  - &  - &  \fullcirc & -0.07 &  - &  - &  - &  - & 310 & 28\\
$M2a2\beta1/2i150$ & Disruptive &  \halfcirc & -7.4 &  - &  - &  - &  - & 150 &  \halfcirc  $\longrightarrow$  \emptycirc & -23 & 44 &  0.98 &  0.71 &  0.02 &  - & -\\%27\\
$M2a9\beta1/2i30$ & Disruptive &  \halfcirc  $\longrightarrow$  \emptycirc & -10 & 110 &  0.98 &  1.9 &  0.09 &  - &  \fullcirc & -0.17 &  - &  - &  - &  - & 130 & -\\%13\\
$M2a9\beta1/2i150$ & Non-disruptive &  \fullcirc & -0.24 & 300 &  0.99 &  3.7 &  0.35 &  - &  \fullcirc & -0.21 &  - &  - &  - &  - & 110 & -\\%21\\
\hline
$M2a6\beta1/2i0$ & Disruptive &  \halfcirc  $\longrightarrow$  \emptycirc & -76 & 77 &  0.98 &  1.5 &  0.05 &  - &  \fullcirc & -0.19 &  - &  - &  - &  - & 190 & -\\%\\25\\
$M2a6\beta1/2i60$ & Disruptive &  \fullcirc  $\longrightarrow$  \emptycirc & -0.67 & 170 &  1.00 &  0.18 &  0.15 &  - &  \fullcirc & -0.14 &  - &  - &  - &  - & 110 & -\\%16\\
$M2a6\beta1/2i120$ & Disruptive &  \fullcirc  $\longrightarrow$  \emptycirc & -0.35 & 2100 &  1.0 &  0.30 &  6.8 &  - &  \fullcirc & -0.20 & 1100 &  1.0 &  3.6 &  2.6 &  - & -\\%21\\
$M2a6\beta1/2i180$ & Non-disruptive &  \fullcirc & -0.41 & 140 &  0.98 &  3.6 &  0.12 &  - &  \fullcirc & -0.34 &  - &  - &  - &  - & 130 & -\\%18\\
\hline
$M2a6\nu45$ & Disruptive &  \fullcirc  $\longrightarrow$  \halfcirc$^{\star}$ & -0.35 & 46 &  0.93 &  3.1 &  0.02 &  - &  \fullcirc & -0.21 &  - &  - &  - &  - & 260 & -\\%18\\
$M2a6\nu90$ & Non-disruptive &  \fullcirc$^{\star}$ & -0.19 & 220 &  0.96 &  9.9 &  0.23 &  - &  \fullcirc & -0.34 &  - &  - &  - &  - & 110 & -\\%23\\
$M2a6\nu135$ & Disruptive &  \halfcirc & -12 & 150 &  0.99 &  2.3 &  0.13 &  - &  \fullcirc & -0.18 &  - &  - &  - &  - & 91 & -\\%19\\
\hline
\hline
$M20a6\beta2i30$ & Non-disruptive &  \fullcirc$^{\star}$ & -0.55 & 87 &  0.62 & 33 &  0.05 &  - &  \fullcirc & -0.42 &  - &  - &  - &  - & 93 & -\\%61\\
$M20a6\beta1i30$ & Non-disruptive &  \fullcirc$^{\star}$ & -0.36 & 220 &  0.88 & 26 &  0.19 &  - &  \fullcirc & -0.59 &  - &  - &  - &  - & 97 & -\\%82\\
$M20a6\beta1/2i30$ & Disruptive &  \emptycirc & - & - &  - &  - &  - &  - &  \fullcirc & -0.93 &  - &  - &  - &  - & 140 & 60\\
$M20a6\beta1/4i30$ & Disruptive &  \emptycirc & - &  - &  - & - &  - &  - &  \fullcirc$^{\star}$ &  0.08 & 280 &  0.86 & 40 &  0.27 &  - & -\\%40\\
$M20a6\beta2i150$ & Non-disruptive &  - &  - & 94 &  0.70 & 28 &  0.06 &  - &  - &  - & 2400 &  0.98 & 53 &  8.5 &  - & 95\\
$M20a6\beta1i150$ & Non-disruptive &  \fullcirc$^{\star}$ & -0.65 & 140 &  0.87 & 18 &  0.09 &  - &  \fullcirc & -0.59 &  - &  - &  - &  - & 48 & -\\%34\\
$M20a6\beta1/2i150$ & Disruptive &  \emptycirc &  - &  - &  - &  - &  - &  - &  \emptycirc &  - &  - &  - &  - &  - &  - & 15\\
$M20a6\beta1/4i150$ & Disruptive &  \fullcirc  $\longrightarrow$  \halfcirc$^{\star}$ & -0.56 & 40 &  0.76 &  9.5 &  0.01 &  - &  \fullcirc & -0.59 &  - &  - &  - &  - & 230 & -\\%230\\
\hline
$M20a2\beta1/2i30$ & Disruptive &  \emptycirc &  - &  - &  - &  - &  - &  - &   \emptycirc & - & - &  - &  - & - &  - & 22\\%170\\
$M20a2\beta1/2i150$ & Disruptive &  \emptycirc &  - &  - &  - &  - &  - &  - &  \emptycirc & - &  - &  - &  - &  - &  - & 77\\
$M20a9\beta1/2i30$ & Disruptive &  \halfcirc & -1.8 &  - &  - &  - &  - & 94 &  \fullcirc$^{\star}$ & -0.41 & 140 &  0.69 & 42 &  0.09 &  - & -\\%67\\
$M20a9\beta1/2i150$ & Disruptive &  \emptycirc &  - &  - &  - &  - &  - &  - &  \halfcirc  $\longrightarrow$  \emptycirc & -9.6 & 53 &  0.98 &  0.98 &  0.02 &  - & -\\%57\\
\hline
$M20a6\beta1/2i0$ & Disruptive &  \halfcirc  $\longrightarrow$  \halfcirc$^{\star}$ & -1.5 & 71 &  0.82 & 13 &  0.04 &  - &  \fullcirc & -0.65 &  - &  - &  - &  - & 89 & -\\%27\\
$M20a6\beta1/2i60$ & Disruptive &  \emptycirc &  - &  - &  - &  - &  - &  - &  \halfcirc & -1.08 &  - &  - &  - &  - & 170 & 77\\
$M20a6\beta1/2i120$ & Disruptive &  \halfcirc & -1.4 &  - &  - &  - &  - & 170 &  \halfcirc  $\longrightarrow$  \halfcirc$^{\star\star}$ & -12 & 27 &  0.74 &  7.0 &  0.01 &  - & -\\%86\\
$M20a6\beta1/2i180$ & Disruptive &  \emptycirc &  - &  - &  - &  - &  - &  - &  \halfcirc  $\longrightarrow$  \halfcirc$^{\star\star}$ & -4.4 & 24 &  0.48 & 13 &  0.01 &  - & -\\%150\\
\hline
\end{tabular}
}
\caption{The outcomes of each model: (\textit{first} column) the model name, (\textit{second} column) the class of encounters, \textit{non-disruptive encounter} or \textit{disruptive encounter} (see \S\ref{subsec:outcome} for their definitions) and (the remaining columns) the information of outcomes and the BH. The columns for the properties of outcomes and the BH present the type (\fullcirc: star whose internal structure remains intact, \halfcirc: partial disruption, \emptycirc: full disruption), $\Delta M/M$ the mass loss in percentile, $a$ the semimajor axis in $\Rsol$, $e$ the eccentricity, $r_{\rm p}$ the pericenter distance in $\Rsol$, $P$ binary orbital period in year and $v$ the ejection velocity at infinity in km/s. $v$ (P) is presented only for unbound (bound) remnants. All the properties of the outcomes presented in the table are measured at the end of the simulation; for the cases where a star forms a bound pair with the BH, we predict their outcomes (circle symbols after $\longrightarrow$) based on their orbit and the disruption (see \S\ref{subsec:outcome} for more detailed descriptions). The superscript ${}^{\star}{}^{\star}$ indicates interacting binaries found before the end of the simulation. Because of the significant mass loss near pericenter, the orbit of the interacting binaries does not settle until the end of the simulation; the orbital elements are measured at the end of the simulation ($t/t_{\rm p}\simeq 80$ for Model $M20a6\beta1/2i(120,180)$ and $\simeq120$ for Model $M20a2\beta1/2i30$). We only include $v$ for the single BH. On the other hand, the subscript $^{\star}$ indicates potential interacting binary candidates, i.e., non-interacting hard binaries (orbital velocity $\gtrsim 50\km/\sec$, c.f., typical velocity dispersion of clusters $\simeq 10-15\km/\sec$ \citealt{Cohen+1983}) (see \S~\ref{sec:implication}). }\label{tab:outcome}
\end{table*}

%\end{sidewaystable}

\begin{figure*}
	\centering
	\includegraphics[width=4.3cm]{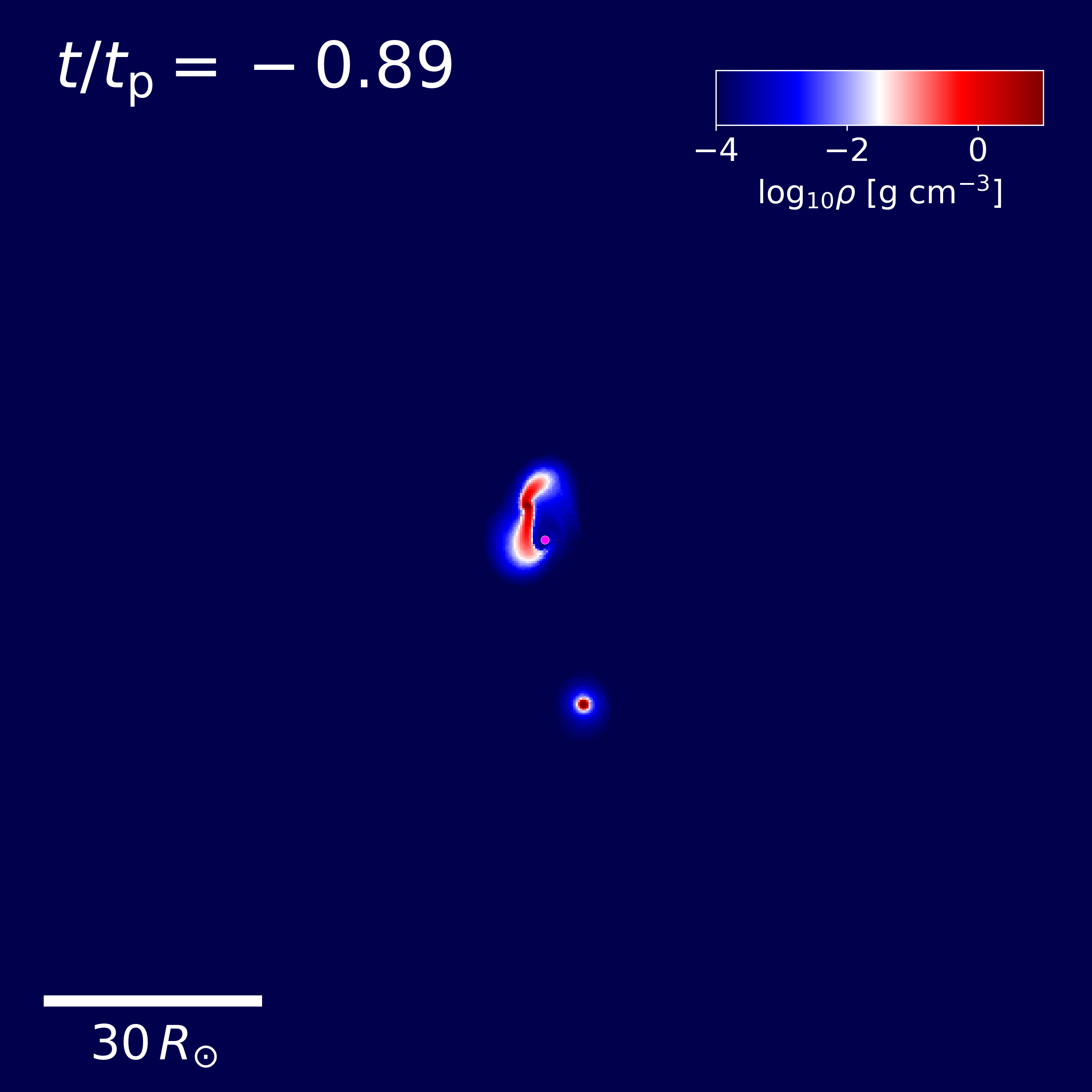}
		\includegraphics[width=4.3cm]{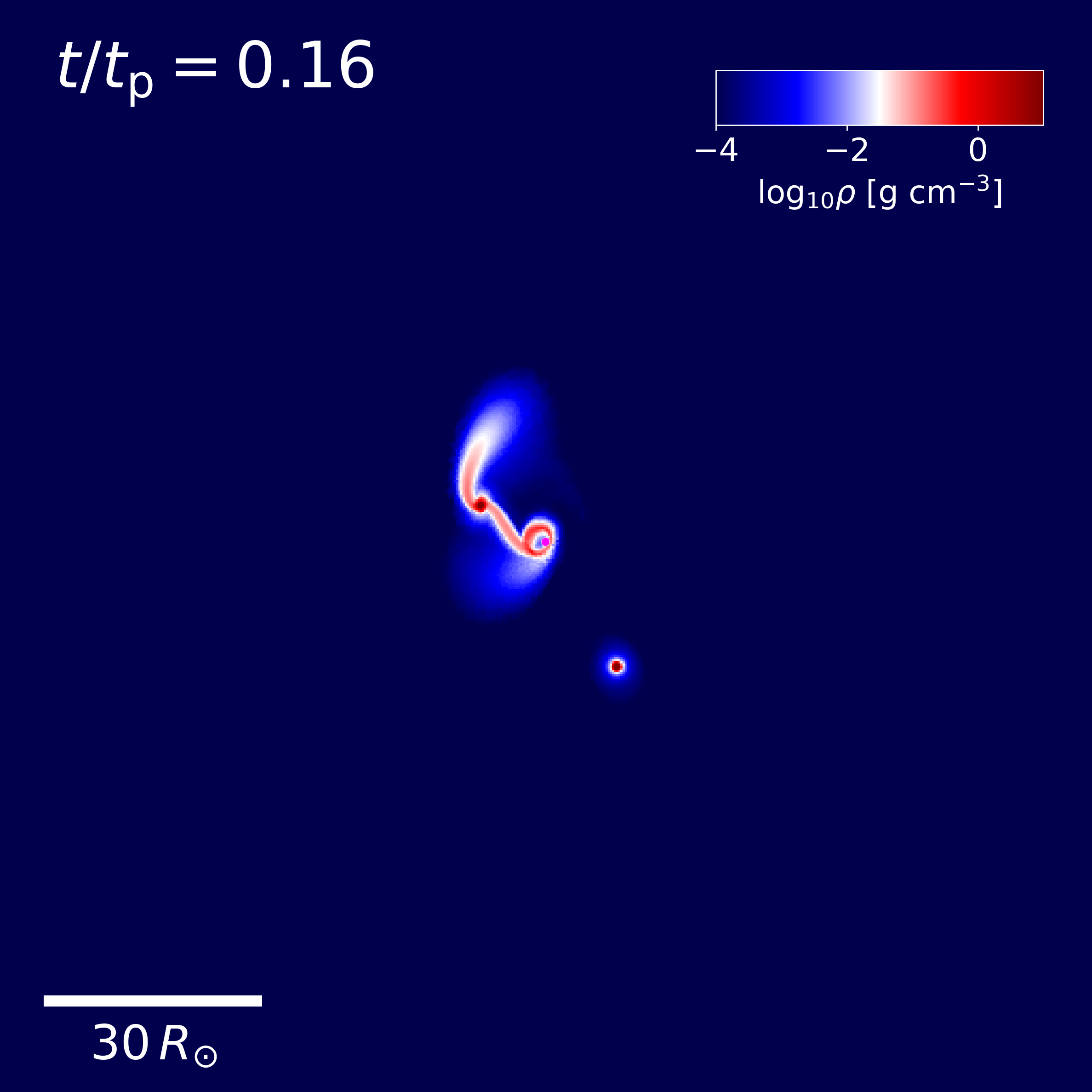}
	\includegraphics[width=4.3cm]{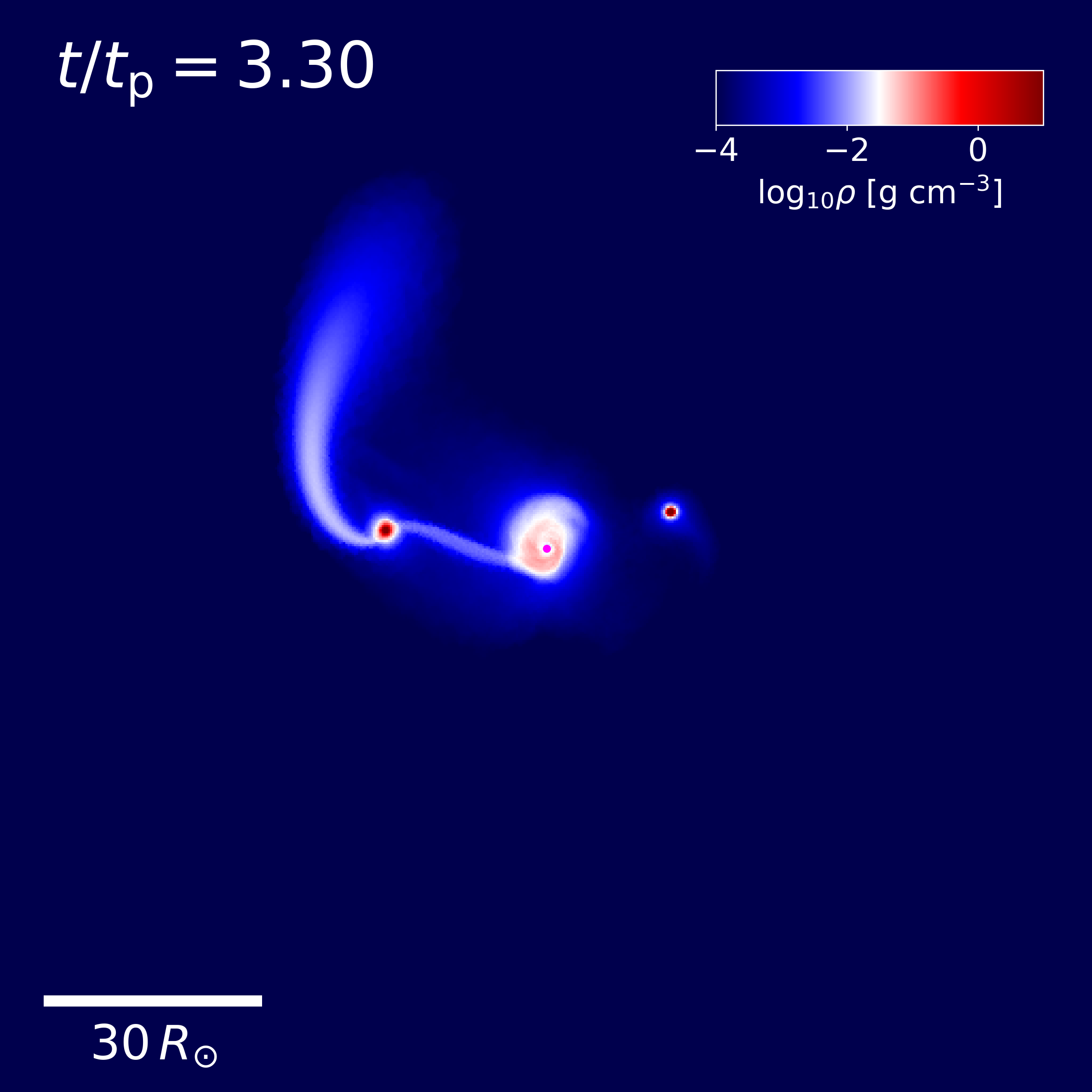}
	\includegraphics[width=4.3cm]{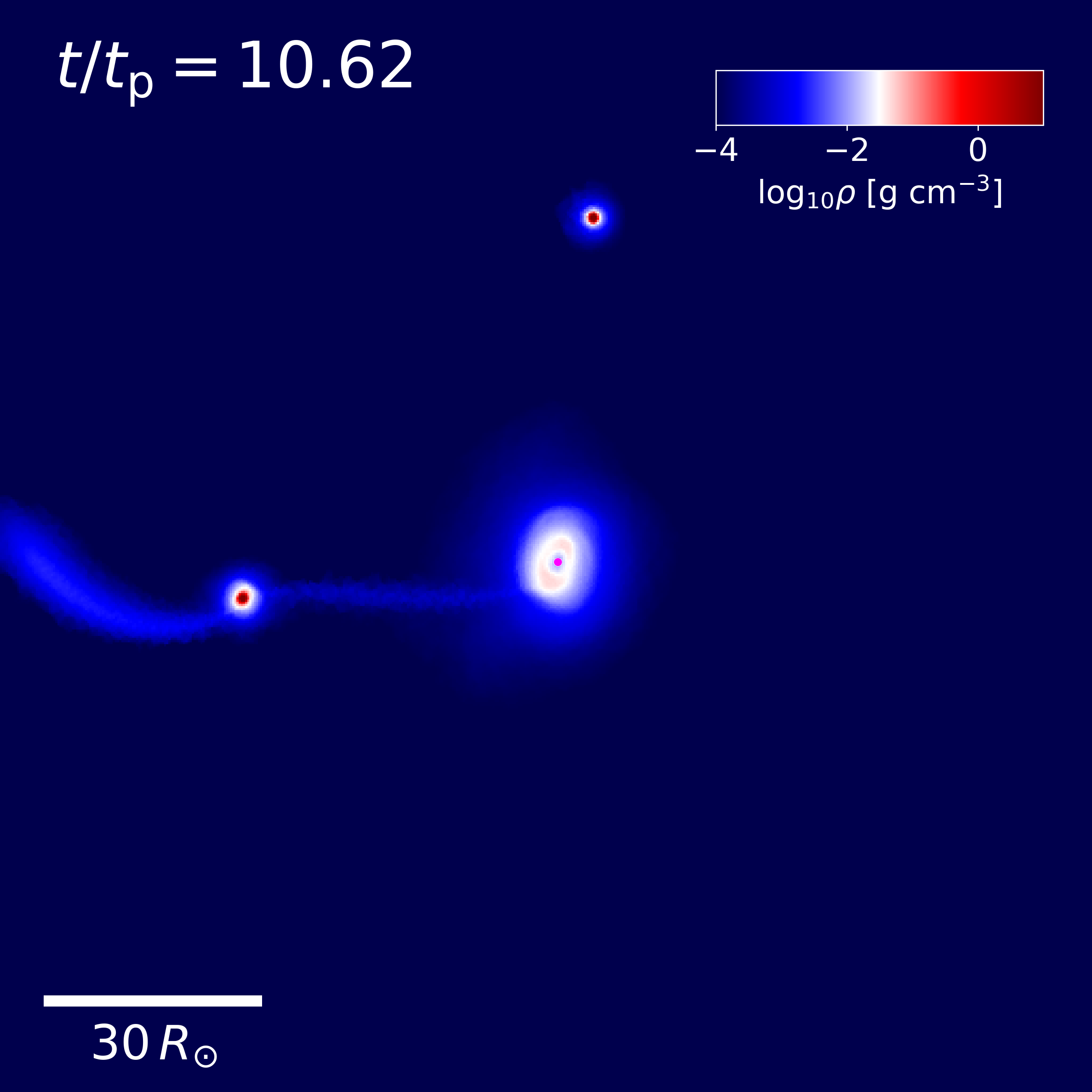}\\
	\includegraphics[width=4.3cm]{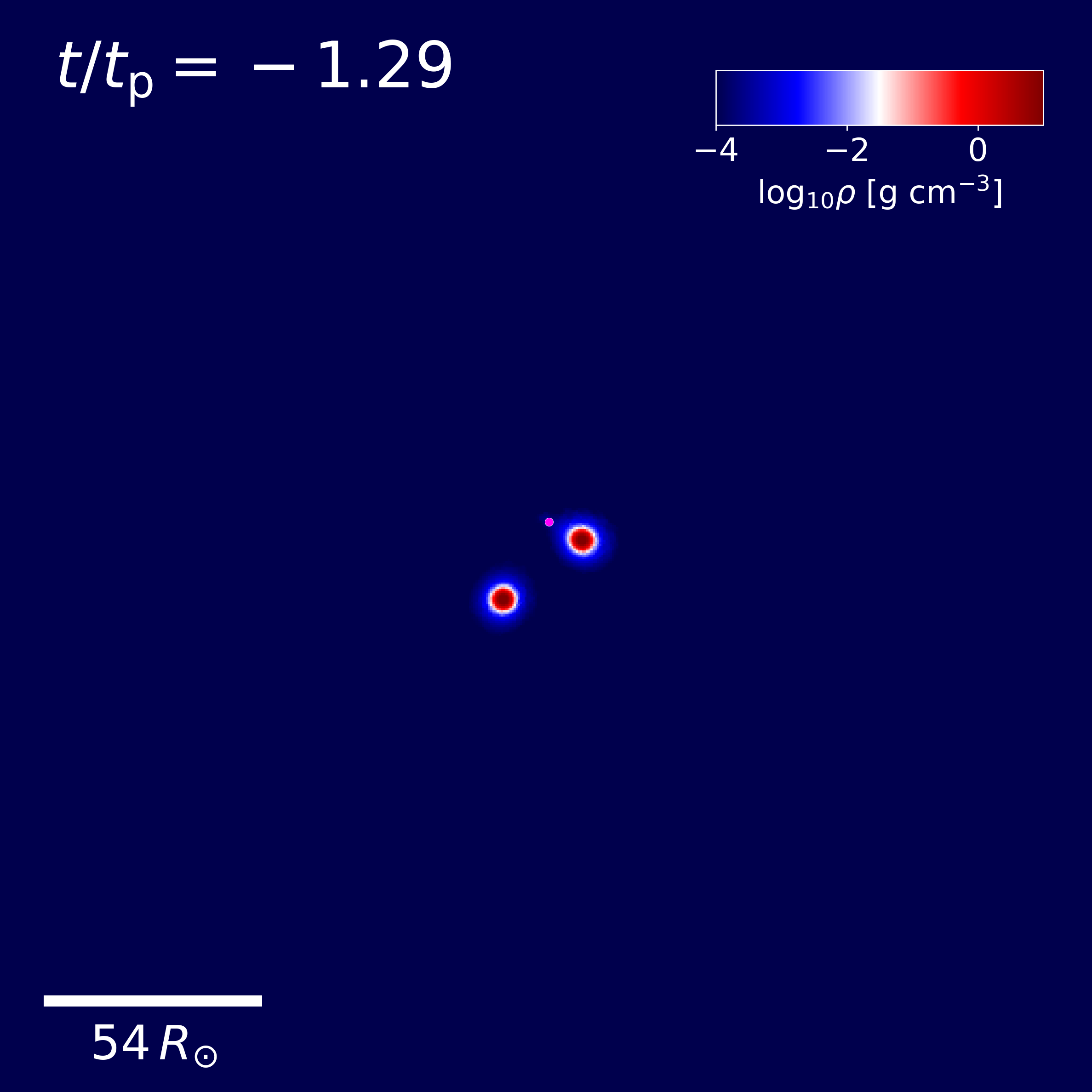}
	\includegraphics[width=4.3cm]{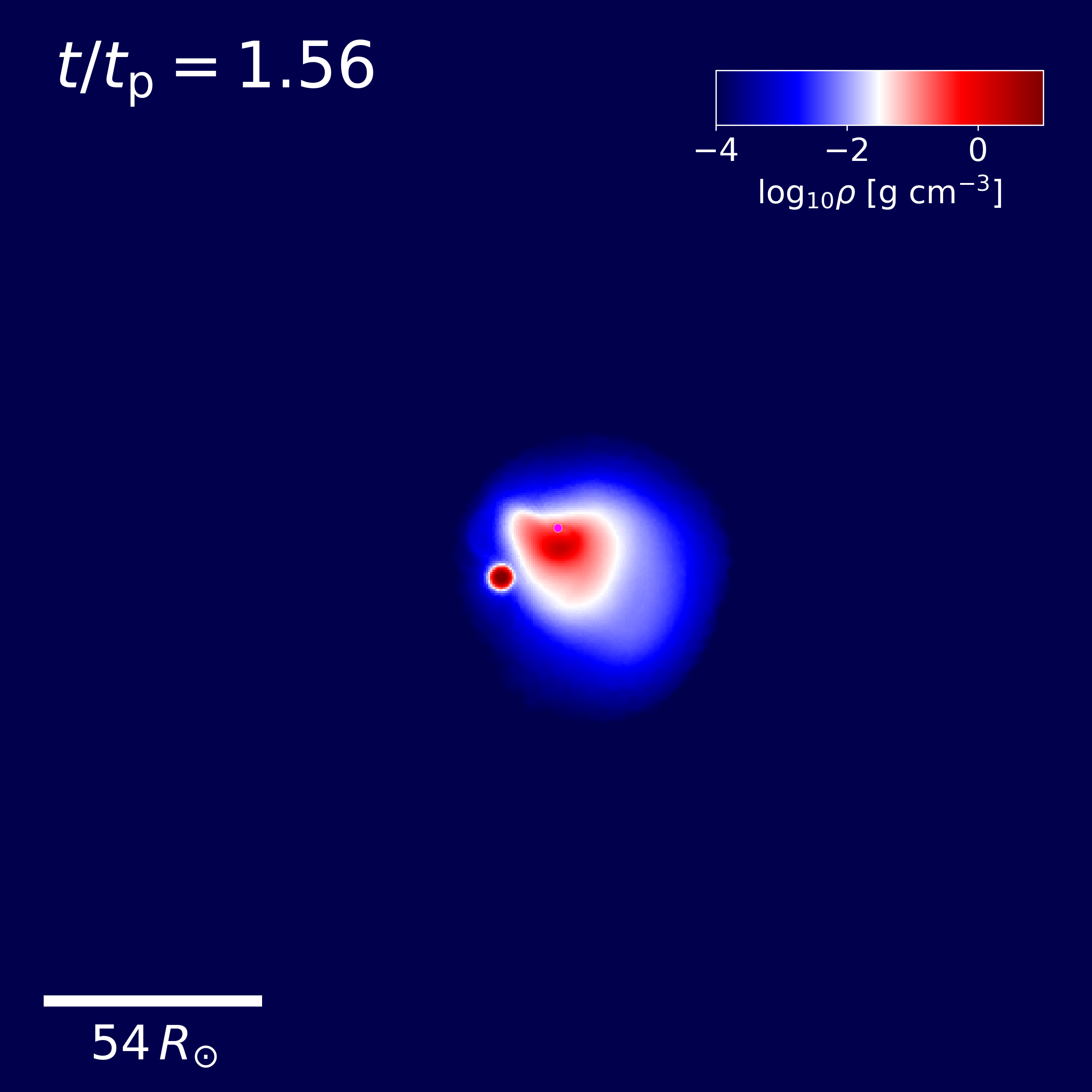}
	\includegraphics[width=4.3cm]{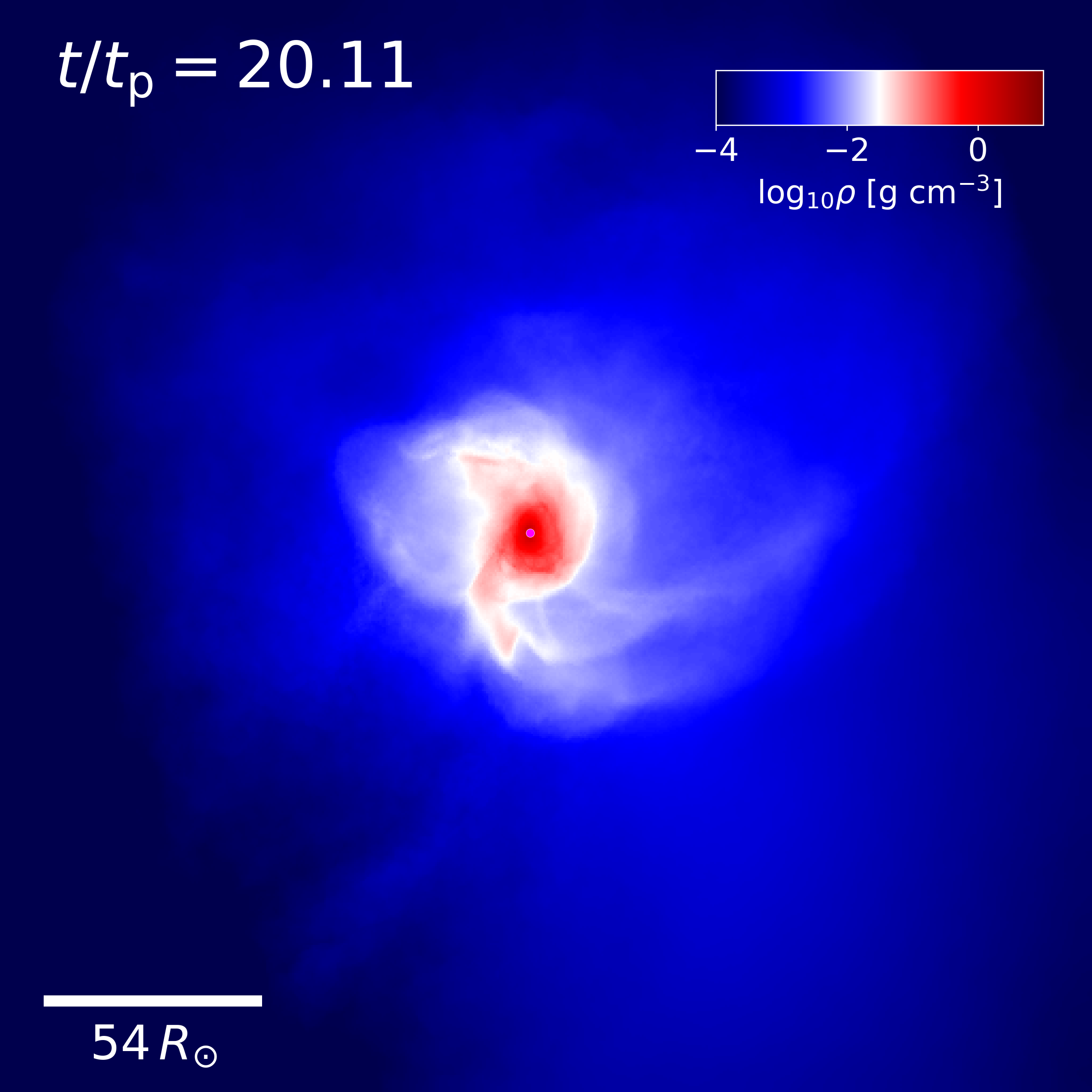}
	\includegraphics[width=4.3cm]{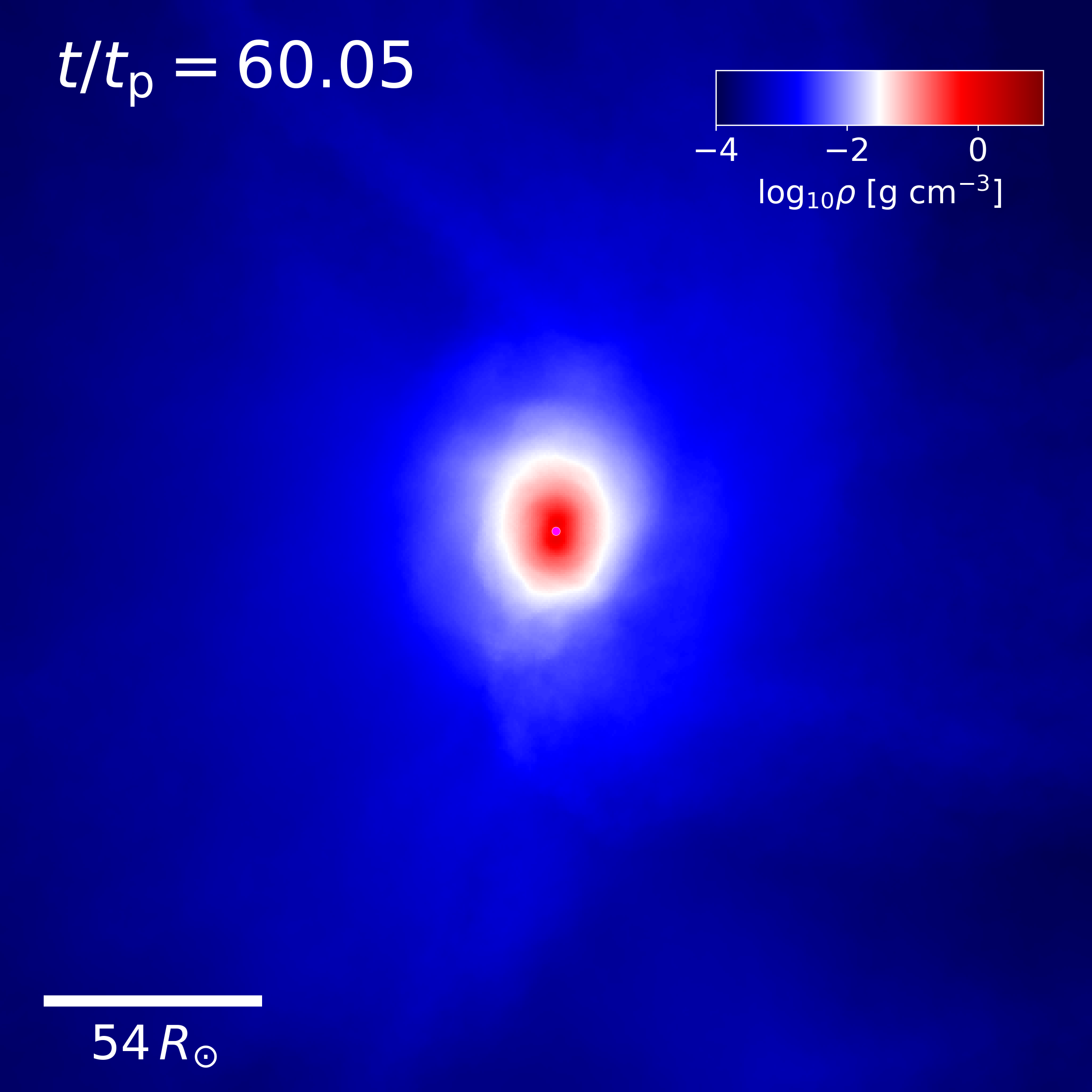}
\caption{The density distribution in the binary orbital plane in two models (\textit{top}: $M2a6\beta1/2i30$ and \textit{bottom}: $M20a2\beta1/2i150$) at a few different times in units of $t_{\rm p}$. The color bar shows the logarithmic density in ${\rm g}/{\rm cm}^{3}$. In the \textit{top} panels, a star is partially disrupted by the black hole (magenta dot) at the first pericenter passage while the other star flies away without being significantly perturbed by the BH. In the \textit{bottom} panels, both stars are disrupted by the BH almost simultaneously at the first contact.}
	\label{fig:example}
\end{figure*}

\section{Results}\label{sec:result}

\subsection{ Classification of outcomes}\label{subsec:outcome}

When the  binary star and the BH encounter, one or both stars can be fully or partially destroyed by the BH, which produces electromagnetic transient (EMT) phenomena. Thus it is important to define those events in a more quantitative way to classify the outcomes. 

We define a full disruption as the case where the star is completely destroyed and no self-gravitating object is left behind. On the other hand, we define a partial disruption as an event where the star loses more than 1\% of its mass and the remnant survives. For other cases where the BH forms a wide and isolated binary whose period is much longer than the simulation duration, we determine the fate of the companion star based on its orbit\footnote{We first compare the orbital period $P_{\rm b-\bullet}$ with the order-of-magnitude estimate of the 1+2 encounter time scale for the binary, $t_{1+2}\simeq [n\Sigma\sigma]^{-1}$ where $n = 10^{5}\pc^{-3}$, $\Sigma$ is the gravitational focusing cross-section and $\sigma = 15\km/\sec$ is the velocity dispersion.  If $P_{\rm b-\bullet}>t_{1+2}$, the star's orbit would be perturbed before it returned to the BH. Although we choose the specific values of $n$ and $\sigma$ relevant for typical clusters, the particular choices do not affect the classification because $t_{1+2}$ is many orders of magnitude greater than the period of the largest binary in our simulations.}. More specifically, to see if the star would undergo another disruption event upon return to the BH, we compare the full (Equation 16 in \citealt{Ryu+2020b}) and partial (Equation 17 in \citealt{Ryu+2020b}) tidal disruption radii  to the pericenter distance\footnote{For the calculation of the partial disruption radius, we define the size of a remnant as the distance from the remnant's center enclosing $99\%$ of its mass. Note that these analytic formulae for the disruption radii are scaled to match the results of the relativistic simulations for TDEs of realistic MS stars by \citet{Ryu+2020a,Ryu+2020b,Ryu+2020c,Ryu+2020d}. This means the expression for the partial disruption radius would give a shorter distance than the actual value considering other effects, such as stellar spin and hotter interior, that would make the remnants more subject to a partial disruption. On the other hand, because the expression for the full disruption radius is determined by the core density, we expect that the full disruption radius would be a more robust estimate than the partial disruption radius.}. If the pericenter distance is longer than the partial disruption radius, the binary would remain a non-interacting one until encountering another intruder. If the pericenter distance is smaller than the partial disruption radius and larger than the full disruption radius, the star would undergo at least one more partial disruption event. Finally, if the pericenter distance is smaller than the full disruption radius, the star would be totally disrupted in one orbital period. 

Applying these definitions to the outcomes at the end of simulations, we can categorize our models into two classes, depending on whether EMT phenomena are created during interactions.
\begin{enumerate}
    \item \textit{Non-disruptive encounter}:  This class corresponds to the case where none of the stars are significantly affected during interactions and survive. So the number of surviving stellar objects is always two. This class happens when the binary and the BH encounter with a large impact parameter, resulting in a perturbation of the binary orbit (without being dissociated) or dissociation of the binary into two single stars. In the case with binary dissociation (Models $M2a6\beta2i150$ and $M20a6\beta2i150$), the semimajor axis of the binary does not change much (less than $1\%$ in $M20a6\beta2i150$) or increases (by a factor of 1.5 in $M2a6\beta2i150$). But the binary becomes eccentric ($e\simeq 0.7$ in both Models). The orbit of the binary relative to the BH becomes a very eccentric ($e_{\rm b-\bullet}\simeq 0.98-0.996$) but more bound ($a_{\rm b-\bullet}\simeq 10-27\AU$ or $P_{\rm b-\bullet}\simeq 8-100\yr$). As a result, the impact parameters of the perturbed binaries' orbits relative to the BH are smaller than those of the initial orbit:  $\beta\simeq 0.67-0.83$ (using the same definition of $r_{\rm t}$ in \S\ref{sub:initialparameter}). 
    
    On the other hand, the case involving binary dissociation usually happens when $1\leq \beta\leq 2$ and $a/a_{\rm RL}=6$ (e.g., Models $M2a6\beta1/2i30$ and $M10a6\beta1/2i30$), with no strong dependence on stellar mass and inclination. In all these cases, one star is bound and the other star is unbound. This is effectively the Hills mechanism \citep{Hills1988} (i.e., the mechanism for the dissociation of a binary by a supermassive black hole, resulting in bound and unbound stars) by a stellar-mass black hole  (or ``micro-Hills mechanism''). 
    The ejection velocity at infinity of the unbound stars is about $50 - 270\km/\sec$. On the other hand, the bound stars are on eccentric or nearly parabolic orbits ($e_{\star-\bullet}\simeq 0.6-1$) with $a_{\star-\bullet}\simeq 0.3-5\AU$ (or $P_{\star-\bullet}\simeq 11$~days to 3 years). Based on the comparison between the tidal radius and the pericenter distance of the bound stars' orbit, there are some cases (e.g., Models $M2a6\beta1i30$ and $M2a6\beta1/4i150$) where the bound stars would undergo at least one more partial disruption event upon return.
    
    \item \textit{Disruptive encounter}: This denotes the case where at least one star is partially or fully disrupted. The most violent case in our simulations associated with disruptions of two stars: full destruction of both stars (e.g., Models $M20a2\beta1/4i30$ and $M20a2\beta1/4i150$) or a full disruption of one star and at least one partial disruption of the other star (e.g., Model $M10a2\beta1/4i30$). As an example, we depict in Figure~\ref{fig:example} the density distribution at four different times in two Models where one star is partially disrupted (e.g. Model $M2a6\beta1/2i30$) and two stars are almost instantaneously disrupted (e.g., Model $M20a2\beta1/2i150$). 
    
\end{enumerate}

This classification is also relevant for EMT formation in multi-body encounters with distinctive observational signatures (see \S\ref{subsec:bhanddebris}). Between the two classes, EMT would definitely be produced in the second class \textit{disruptive encounter}. Nonetheless, there is still the possibility of the formation of EMT in the first class \textit{non-disruptive encounter}: the micro-Hills mechanism produces a hard binary (e.g., SP$^{\star}$ in Table~\ref{tab:outcome}). If the orbit continues to shrink via weak encounters with other star, the BH will generate EM flares by accretion of overflowing gas from the star ($\S\ref{subsec:X-raybinary}$).

We summarize the classification of each model and the properties of the outcomes in Table~\ref{tab:outcome}.

\begin{figure}
	\centering
	\includegraphics[width=8.0cm]{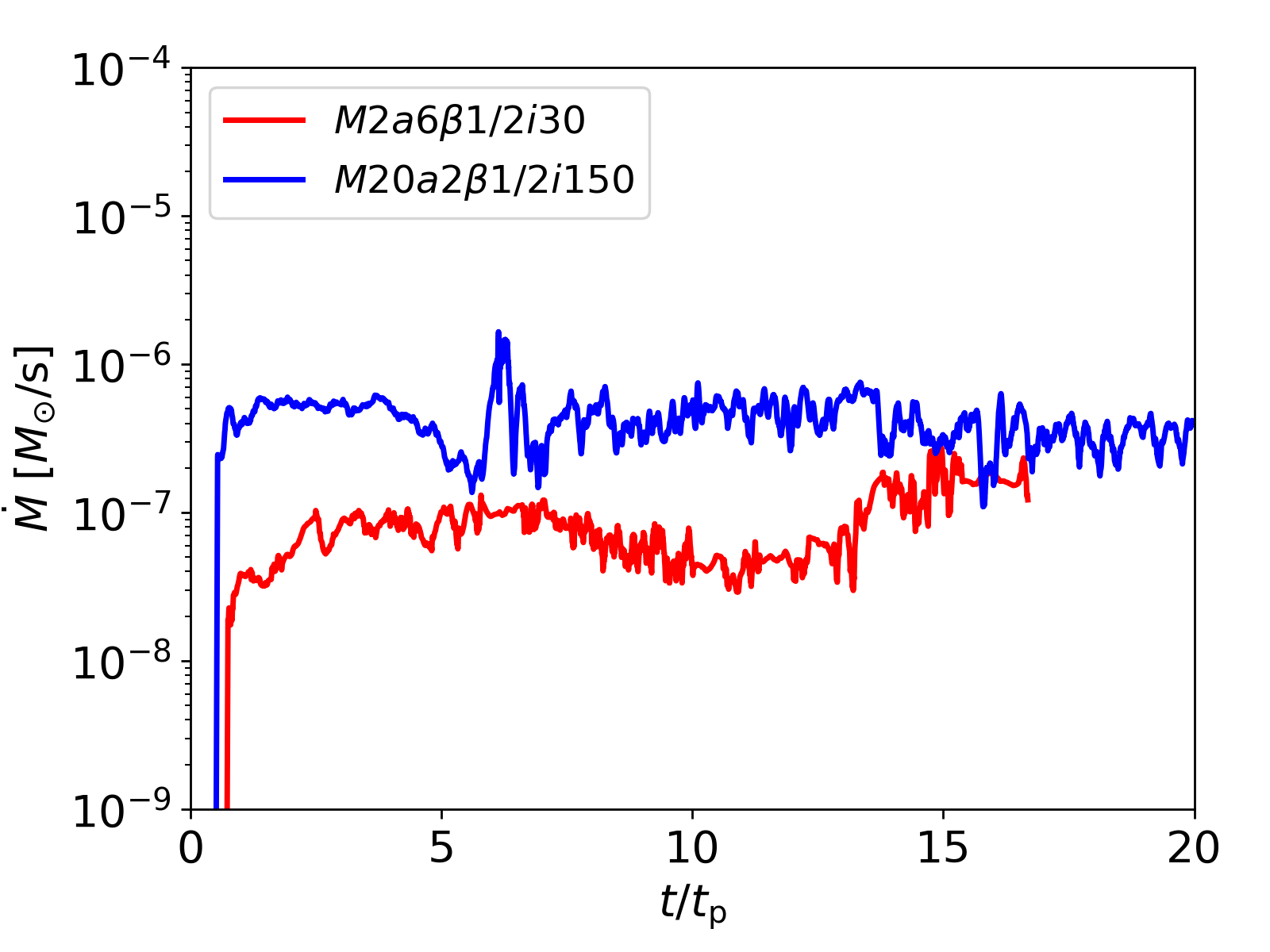}
\caption{The accretion rate $\dot{M}$ for the two models, $M2a6\beta1/2i30$ (red) and $M20a2\beta1/2i150$ (blue), as a function of $t/t_{\rm p}$ since the first closest approach. Here, $t_{\rm p}\simeq 2.2$ (1.2) hours for Model $M2a6\beta1/2i30$ ($M20a2\beta1/2i150$). }
	\label{fig:mdot}
\end{figure}

\subsection{Dependence of outcomes on encounter parameters}\label{subsec:dependence}
In this study, we examine the dependence of outcomes on a few key encounter parameters:  $M_{\rm b}$ (mass of the initial binary), $a$ (semimajor axis of the initial binary), $\beta$ (impact parameter), $i$ (inclination angle) and $\nu$ (phase angle by varying one parameter at a time while the rest are held fixed. Although our models do not cover the entire parameter space, we can see some clear trends. Our simulations suggest that there are three main parameters which determine the location of the boundary between \textit{non-disruptive} interactions and \textit{disruptive} interactions, namely, the impact parameter $\beta$ and binary size $a/a_{\rm RL}$ and the phase angle $\nu$. 

\begin{enumerate}
    \item \textit{Dependence on the impact parameter $\beta$}: as one can see from Models $M(2,20)a6\beta(1-1/4)i(30,150)$ in Tables~\ref{tab:outcome}, three-body encounters with $\beta >1$ are \textit{non-disruptive} interactions, whereas those with $\beta\leq1$ tend to disrupt at least one star. It is not surprising given that $\beta$ directly determines how close the binary and the BH can approach one another. However, we note that a small impact parameter does not always result in disruption events (e.g., Model $M2a6\beta1i150$ and possibly for a very wide binary, see below). In other words, it is not a sufficient condition but a necessary condition for a disruption event, while a large impact parameter is a sufficient condition for \textit{non-disruptive} encounters.

    \item \textit{Dependence on the semimajor axis $a$}: the comparison between Models with different $a/a_{\rm RL}$ (e.g., $M2a(2-9)\beta1/2i30$) suggests that encounters become more disruptive for smaller $a/a_{\rm RL}$ as long as $\beta$ is  sufficiently small. It is because the interactions move away from the regime of two 1+1 encounters towards the regime of chaotic 2+1 encounter. Stars in large binaries can also be disrupted by the BH (e.g., $M(2,20)a9\beta1/2i(30,150)$), but such events become increasingly more like an ordinary TDE by single BH. Interestingly, we found a merger of two partially disrupted stars as an intermediate outcome in one of our models with the smallest $a/a_{\rm RL}$ considered (i.e., Model $M20a2\beta1/2i30$), which is in line with stellar mergers during three-body encounters between stars found in \citet{McMillan+1991}. Note that the semimajor axis of the initial binary in \citet{McMillan+1991} is $4$ and $8R_{\star}$ (cf., $a=5.3R_{\star}$ for our smallest binary). 

Another consideration related to the dependence of $a$ and $\beta$ for wide binaries is that if the impact parameter for encounters between a wide binary and a black hole is so small that the BH's crossing time across the binary is shorter than the binary period (i.e., the binary is ``frozen'' while the BH passes through) and the BH's gravitational force on any of the stars is weaker than the stars' gravitational pull to each other, it may be possible that the BH simply penetrates through the binary in the middle without interacting significantly with any of the stars. This consideration supports that a small impact parameter is a necessary condition for \textit{disruptive} encounters. 

\item \textit{Dependence on the phase angle $\nu$}: even for a sufficiently small impact parameter and small $a$, the outcomes can vary depending on the phase angle, as shown in Models with different $\nu$ (Models $M2a6\beta1/2i30$ and $M2a6\nu(45-135)$). Various outcomes of encounters between a single star and a binary black hole for different phase angles was also reported in \citetalias{Ryu+2022}.  The statistical likelihood of disruptions are mostly governed by $\beta$ and $a$, but it is very important to consider the phase angle for the outcome of an individual encounter case.

\item \textit{Dependence on the inclination angle $i$}:  we found no significant dependence on $i$. The weak dependence on $i$ is at odds with the strong dependence for encounters between a single star and a binary BH found in \citetalias{Ryu+2022}. The comparison is not straightforward given different encounter parameters. However, how chaotic the interactions are (or whether the initial binary is dissociated during encounters) may result in different levels of $i-$dependence: in encounters between a single star and a BH binary considered in \citetalias{Ryu+2022}, the binary is never dissociated by the star at the first closest approach and the star interacts with the binary mostly once before its disruption or ejection. For that case, the relative velocity between the star and the interacting BH at the first closest approach plays an important role in determining the outcome because that directly governs how long the star is tidally affected by the BH. And the relative velocity is different depending on whether the encounter is in a prograde or retrograde direction for given parameters. 
On the other hand, in the simulations of this study, the binary is always dissociated (except for the two models with $\beta=2$), followed by more chaotic interactions, which likely removes the memory of the initial approach. 

\item \textit{Dependence on the binary mass $M_{\rm b}$}: The dependence of $M_{\rm b}$ on the types of outcomes appears to be weak. However, it actually means that the mass (and thus energy) budget for radiation from debris is greater for the disruption of more massive stars (i.e., similar $\Delta M/M$ for larger $M_{\star}$). Furthermore, the momentum kick given to the BH is greater for encounters with more massive binaries. In particular, the ejection velocity of the BH is $\sim 10-30\km/\sec$ in encounters with the $2\Msol$ binary, whereas it is in the range $\sim 50 - 200\km/\sec$ in encounters with the massive stars of our simulations. Note that the ejection velocity of the unbound stars from the $2\Msol$ binary tends to be greater than that for more massive unbound stars, but the degree is not as significant as the ejection velocity of the BH. 

\end{enumerate}

\subsection{Impact on the black hole properties}\label{subsec:bhanddebris}

The immediate impact of close encounters with binary stars on the BH is momentum kick, growth in mass and, therefore, spin evolution. 

The ejection velocity of the BH in most cases is a few tens of km/sec, but there are a few cases with an ejection velocity of $\sim100-200\km/\sec$ (see Table ~\ref{tab:outcome}). Those velocities are either larger than or comparable to the escape velocity of globular clusters (i.e., a few to 180 km/sec; \citealt{Antonini+2016,Gnedin+2002}). We find that the only significant correlation of the ejection velocity is with the mass of the binary that encounters the BH, as discussed in \S~\ref{subsec:dependence}: encounters with a more massive binary results in higher ejection velocity of the BH.

If a star is disrupted at least once during encounters, the ejected BH is surrounded by an accretion flow made from stellar debris, as shown in Figure~\ref{fig:example}. The gas accretes onto the BH and the BH grows in mass and spins up. Typically, the accretion rate remains nearly the same at super-Eddington $\dot{M}\simeq 10^{-8}-10^{-6}\Msol/\sec$ once an accretion disk forms, which is many orders of magnitude higher than typical Roche Lobe overflow rates in interacting BH binary systems \citep{Savonije+1978}. This trend is illustrated in Figure~\ref{fig:mdot} showing the time evolution of the accretion rate for two models, one \textit{single disruption} case (Model $M2a6\beta1/2i30$) and one \textit{double disruption} case (Model $M20a2\beta1/2i150$). Note that these are  the same models shown in Figure~\ref{fig:example}.

\begin{figure*}
	\centering
	\includegraphics[width=4.3cm]{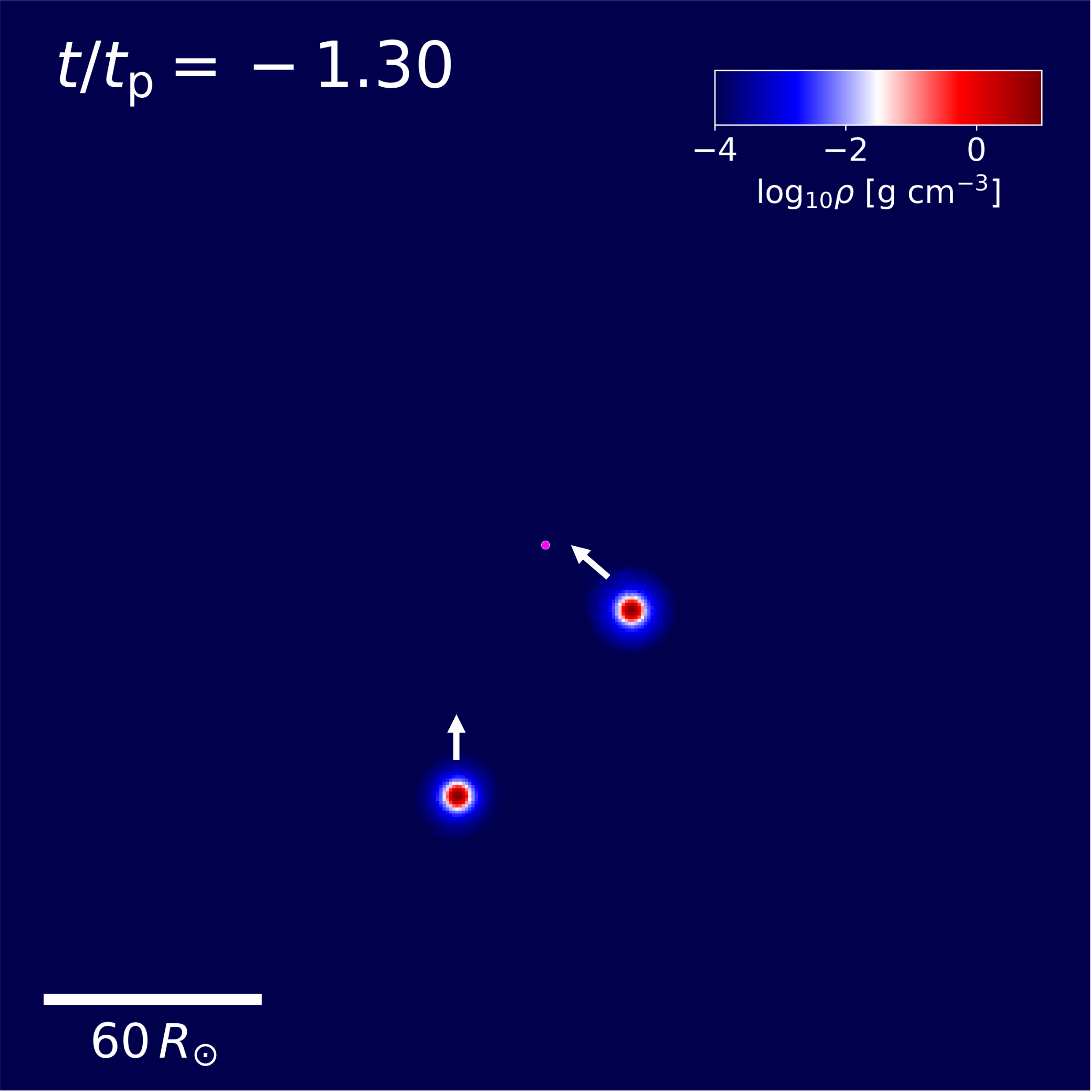}
	\includegraphics[width=4.3cm]{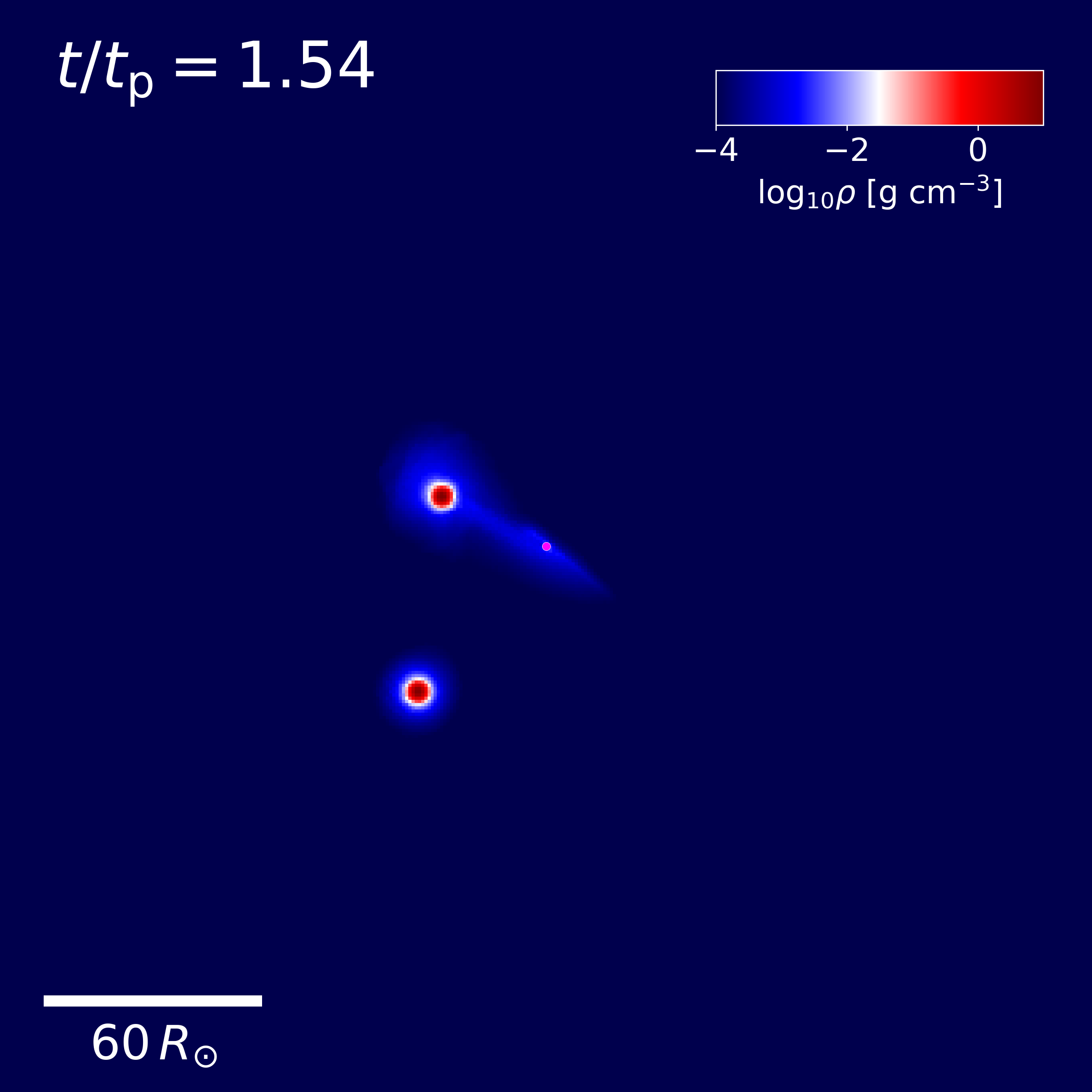}
	\includegraphics[width=4.3cm]{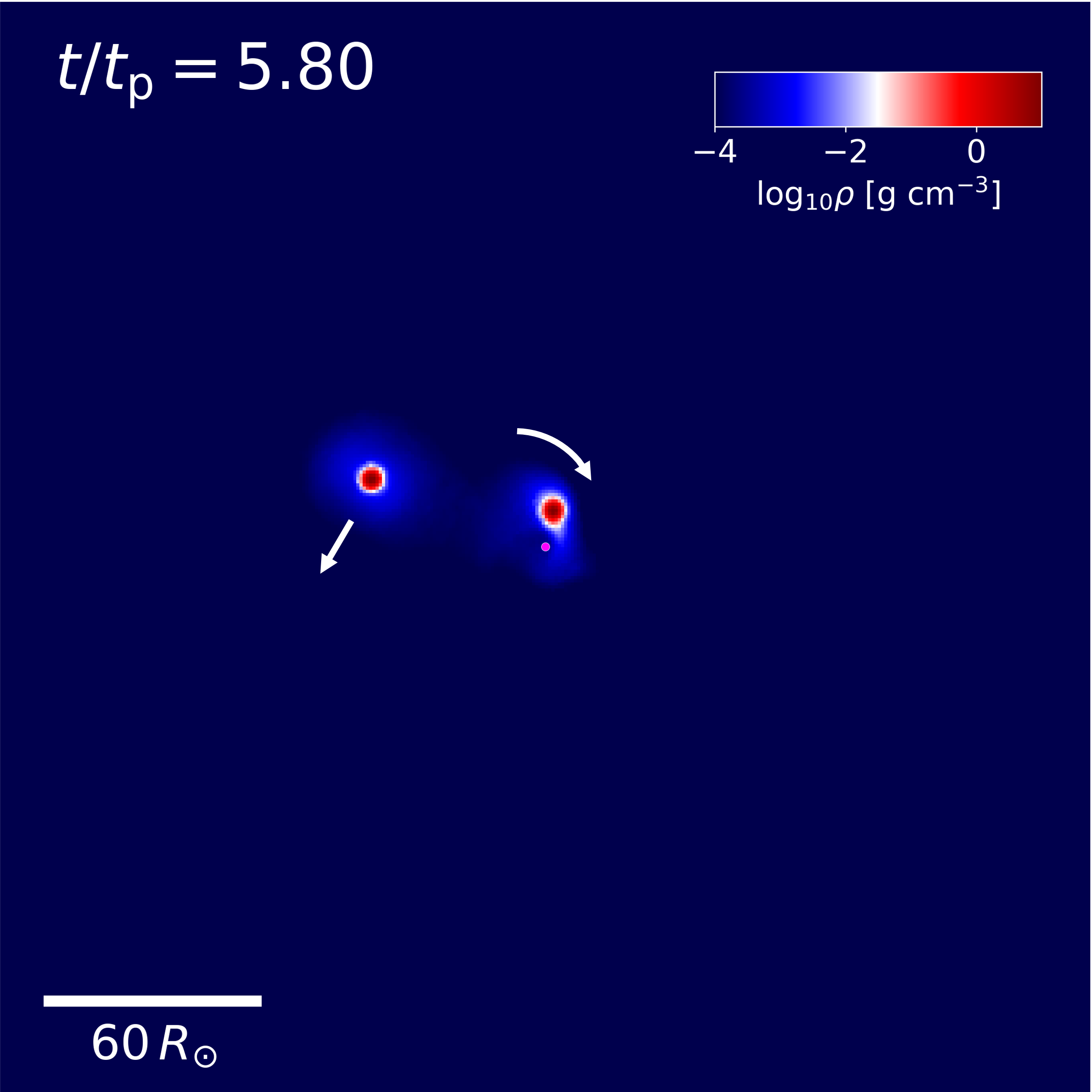}
	\includegraphics[width=4.3cm]{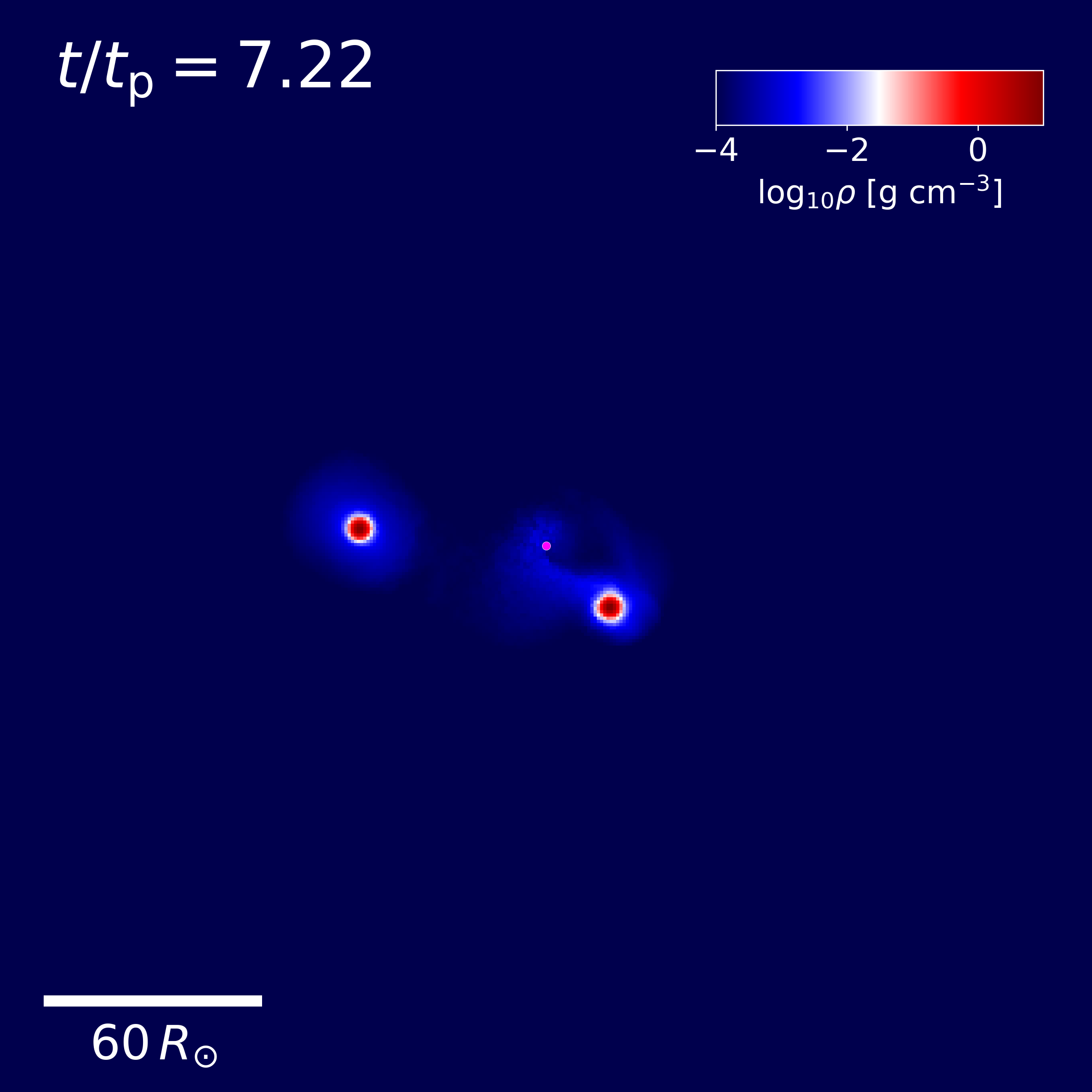}\\
	\includegraphics[width=4.3cm]{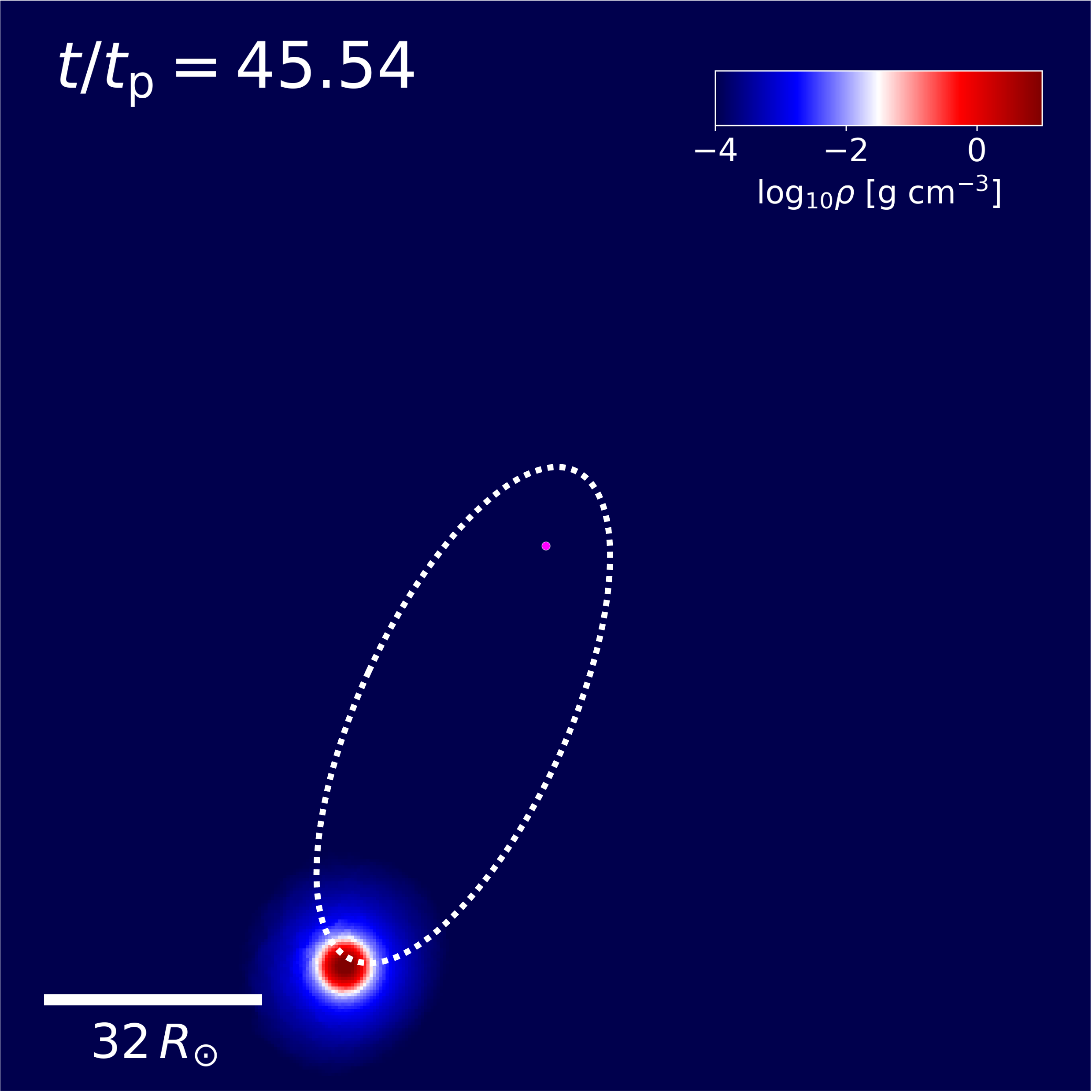}
	\includegraphics[width=4.3cm]{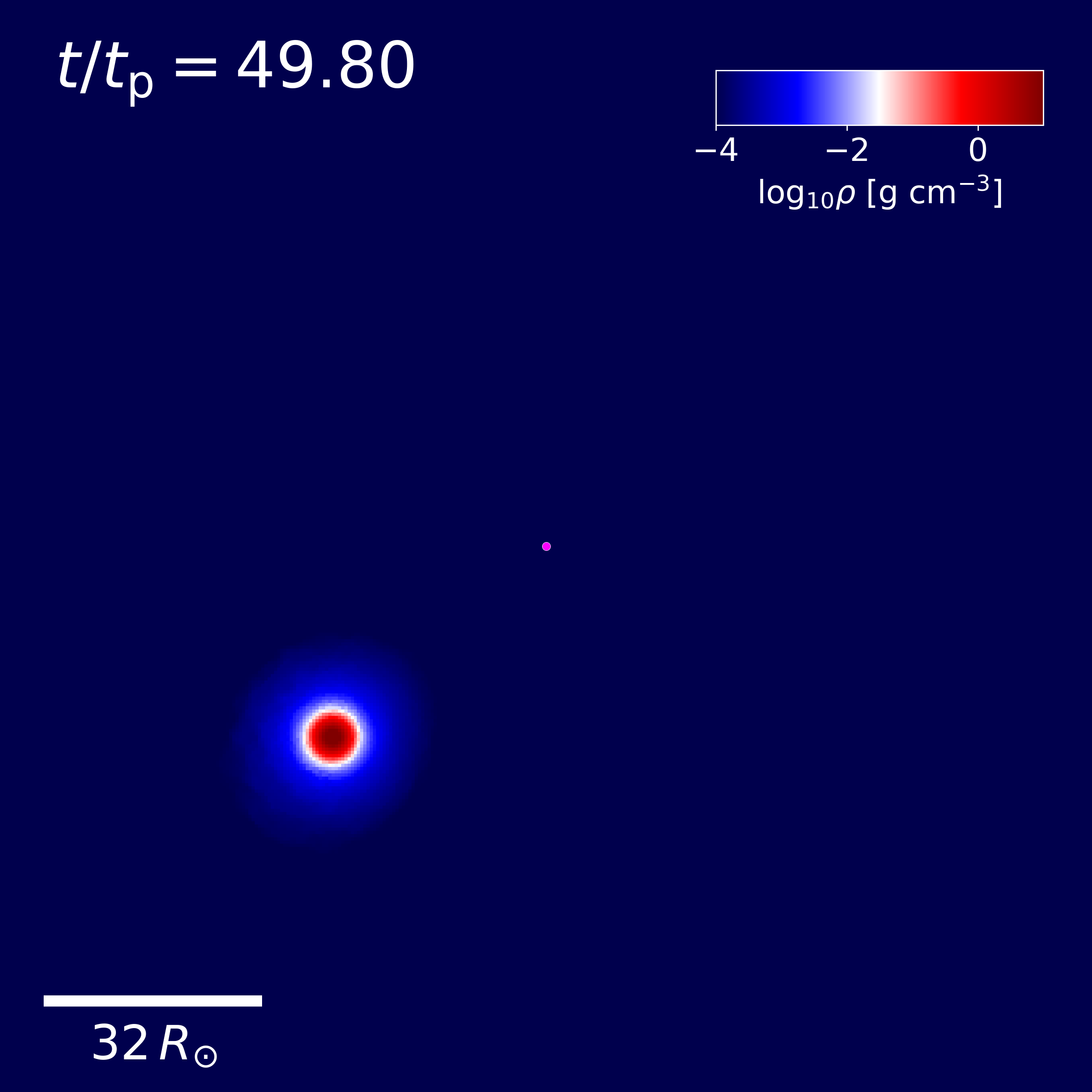}
	\includegraphics[width=4.3cm]{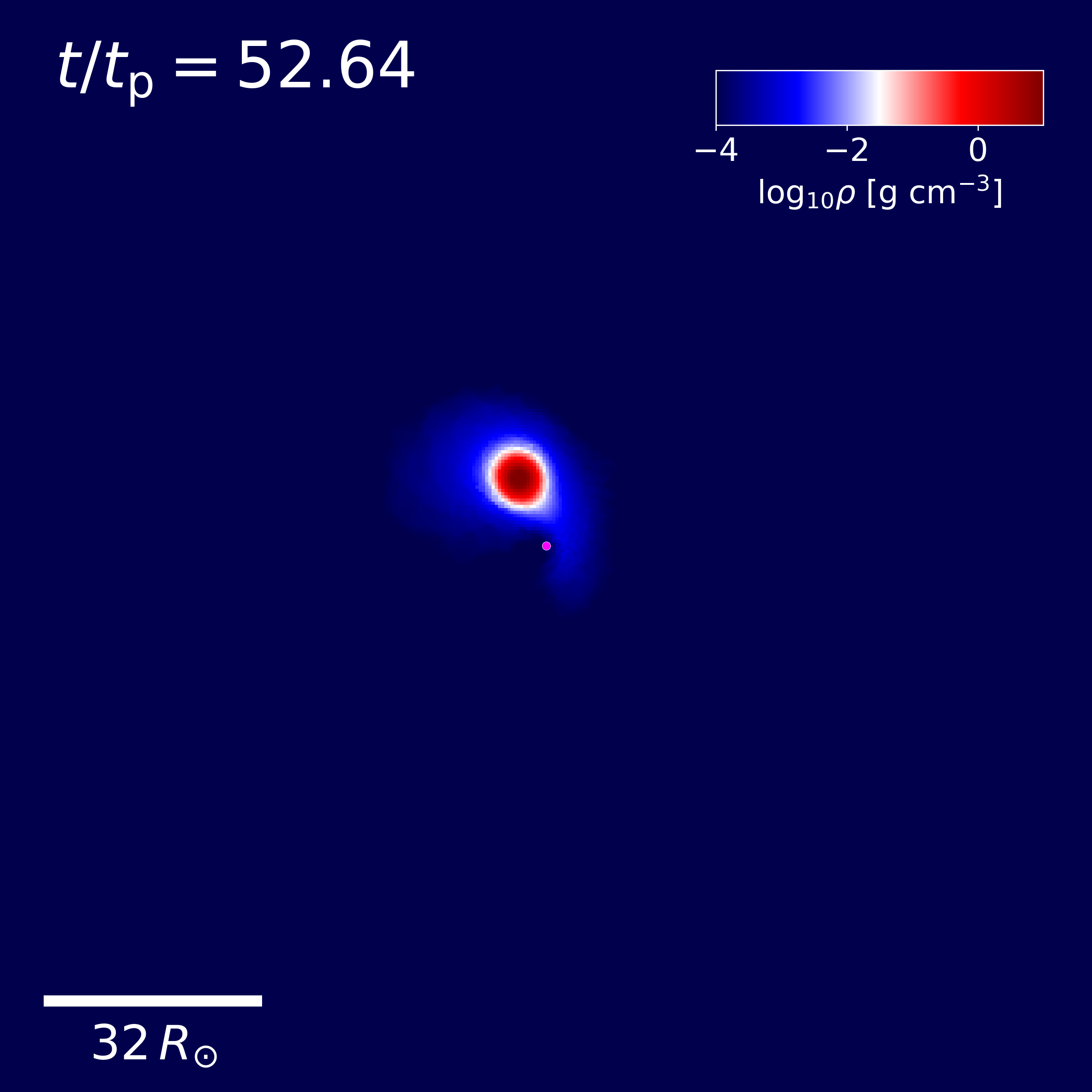}
	\includegraphics[width=4.3cm]{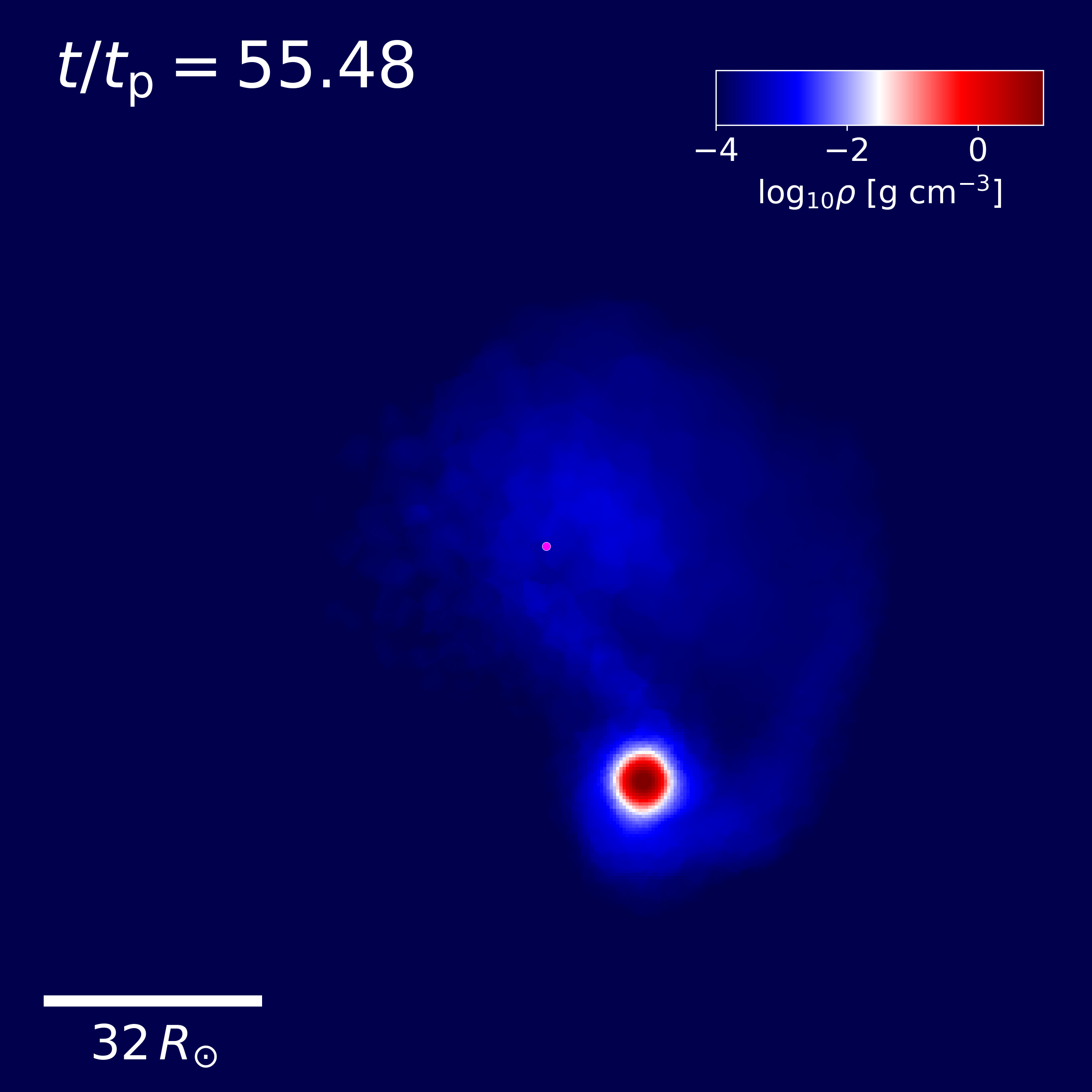}
\caption{Series of snapshots, centered at the BH, for Model $M20a6\beta1/2i120$ showing a star-BH binary at a few different times in units of $t_{\rm p}\simeq 2.2$ hours. The color bar shows the logarithmic density projected in the $x-y$ plane in ${\rm g}/{\rm cm}^{3}$.  In the \textit{lower-left} panel, the dotted eccentric circle approximately shows the orbital trajectory of the star relative to the BH.}
	\label{fig:x_ray_binary}
\end{figure*}

\begin{figure}
	\centering
	\includegraphics[width=8.0cm]{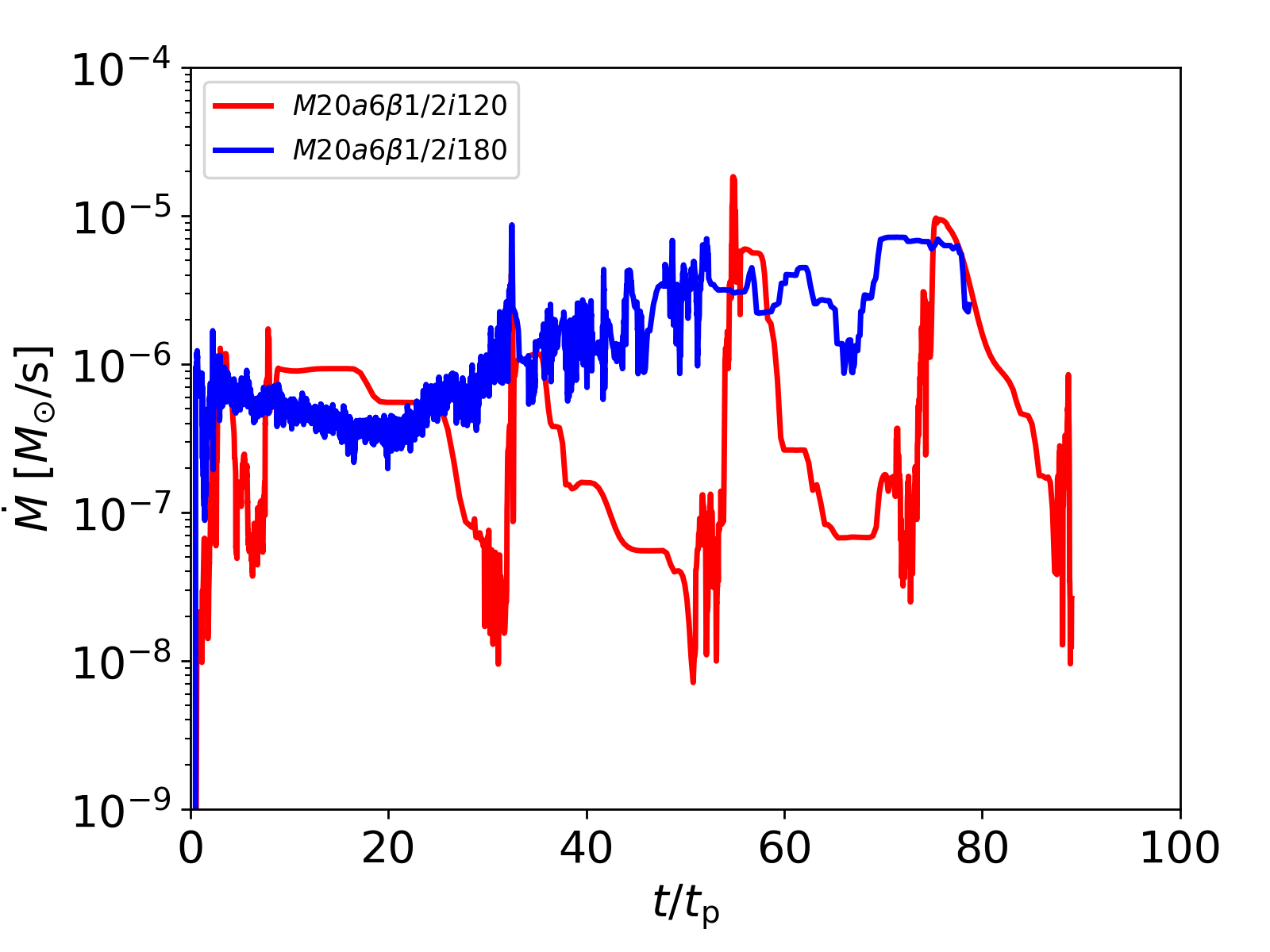}
\caption{The accretion rate $\dot{M}$ (\textit{top}) for interacting binaries found in two Models $M20a6\beta1/2i120$ (red) and $M20a6\beta1/2i180$ (blue). The $t_{\rm p}\simeq 6.4$ hours is the dynamical time at the pericenter distance of the orbit of the initial binary relative to the BH.}
	\label{fig:mdot_Xraybinary}
\end{figure}

\begin{figure}
	\centering
	\includegraphics[width=8.0cm]{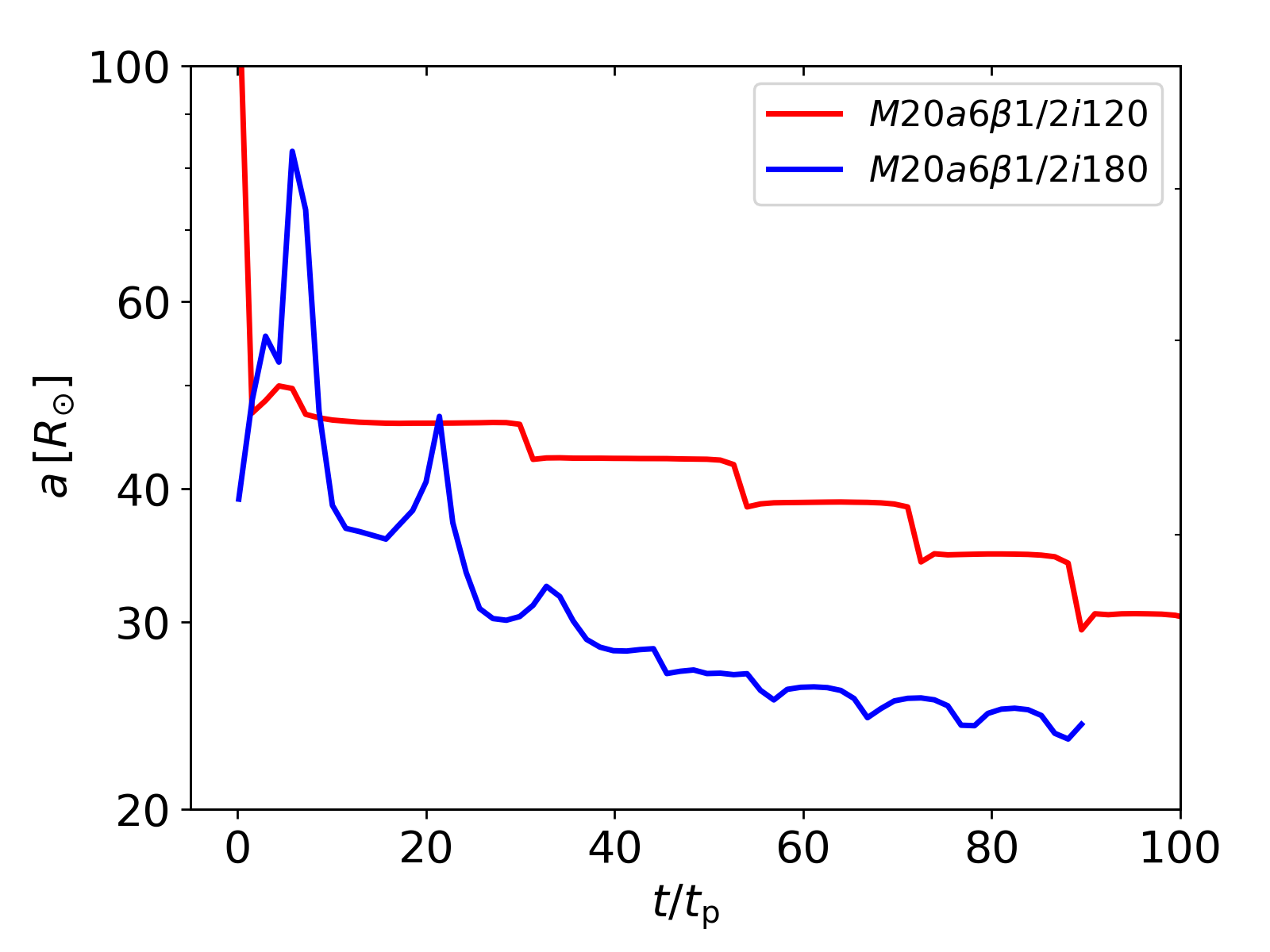}
	\includegraphics[width=8.0cm]{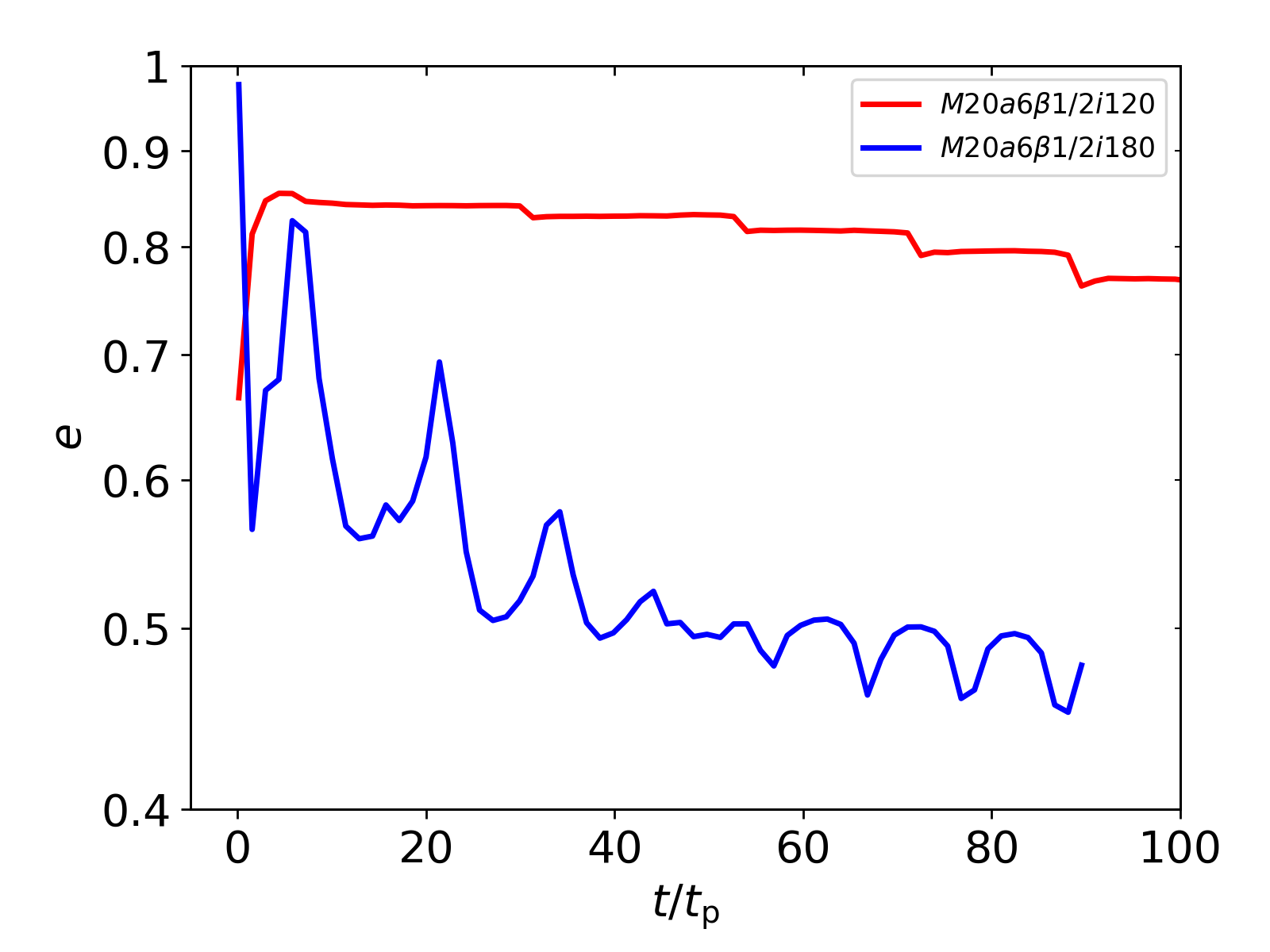}
	\includegraphics[width=8.0cm]{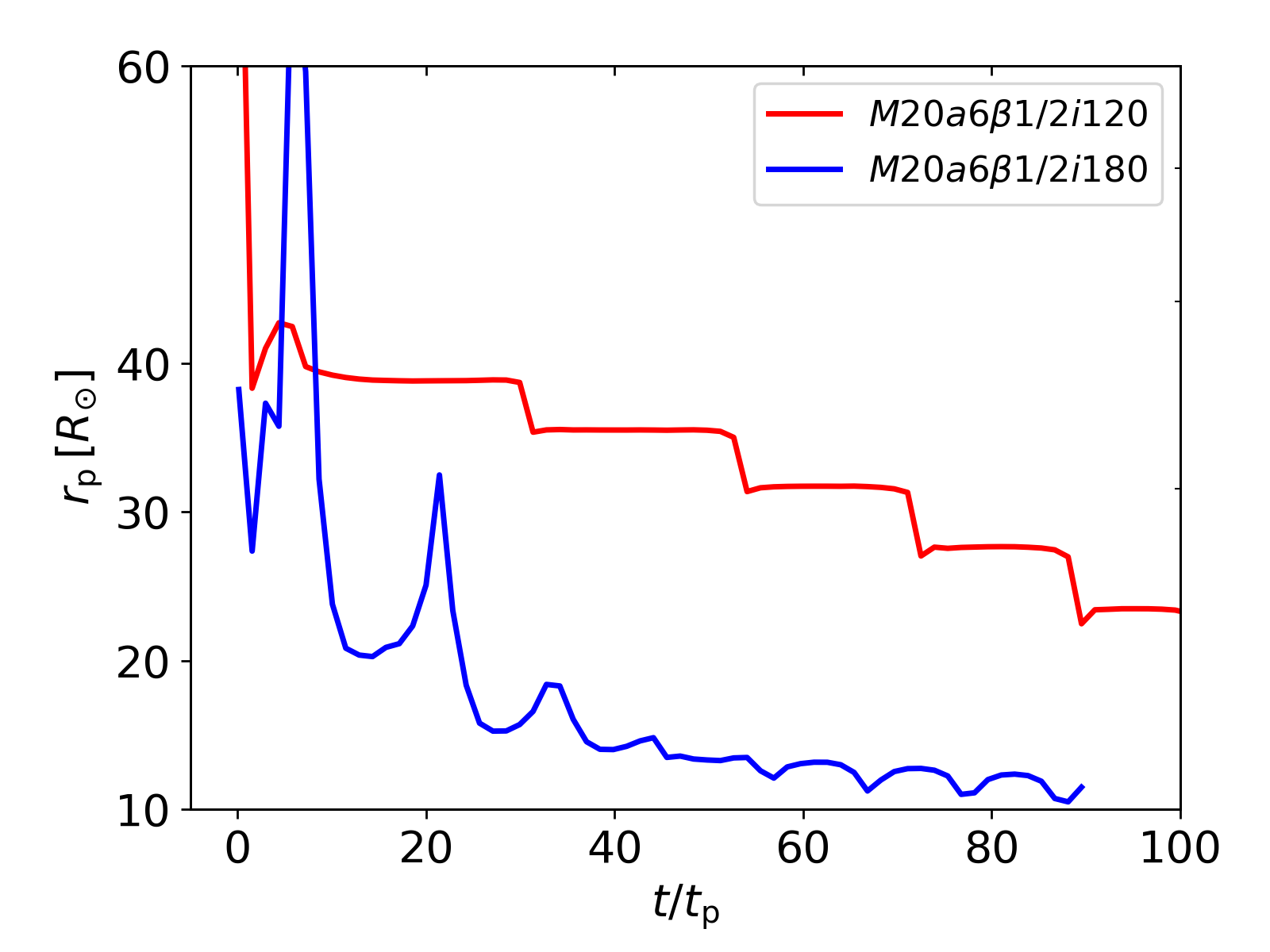}
\caption{The orbital perimeters, semimajor axis (\textit{top}), eccentricity (\textit{middle}) and pericenter distance (\textit{bottom}) for interacting binaries found in two Models $M20a6\beta1/2i120$ (red) and $M20a6\beta1/2i180$ (blue). $t_{\rm p}\simeq 6.4$ hours is the dynamical time at the pericenter distance of the orbit of the initial binary relative to the BH.}
	\label{fig:orbit_Xraybinary}
\end{figure}

\section{Discussion}\label{sec:discussion}

\subsection{Astrophysical Implications}\label{sec:implication}

\subsubsection{Varieties of transient phenomena}

We have reported that dynamical interactions between binary stars and a single black hole can produce a variety of EMT phenomena: a single partial disruption event, a full disruption event, multiple partial, or full disruption events. Even multiple disruption events can be sub-divided into two classes: almost instantaneous double disruption events and successive partial disruption events of bound remnants. Furthermore, an interacting star-BH binary (see \S~\ref{subsec:X-raybinary}) can form while the other star is fully disrupted (e.g., Model $M20a6\beta1/2i120$) or the other star is ejected via the Hills mechanism \citep{Hills1988} (or ``micro-Hills'' mechanism). Dynamical interactions can create ``collision''-like disruption events where disk formation is prompt. Such varieties constitute the most significant difference from the ordinary TDEs, which are disruptions of a single star by a single black hole.  In particular, double disruption events can never happen in ordinary TDEs.

Such varieties of EMT phenomena and the chaotic nature of three-body interactions indicate that light curves of TDEs or almost collision-like disruptions in three-body interactions should look qualitatively different from those of ordinary TDEs. Detailed quantitative predictions of light curves are beyond the scope of this paper and reserved to future work. Nonetheless, we can still make qualitative predictions for the expected observational signatures of double disruption events. For the case of rapid double full disruption, almost instantaneous disk formation may result in a sudden increase in luminosity to super-Eddington in EUV and X-rays (luminosity ($\propto T^{4}$)-weighted average of temperature $\simeq 10^{6}-10^{7}$K), which, if the luminosity $\propto \dot{M}$, may remain roughly constant up to a few months ($\simeq O(1\Msol)/O(10^{-6}\Msol/\sec)$).

Furthermore, the total energy budget ($\propto M_{\star}$) is greater than for a single full disruption. However, if super-Eddington accretion causes significant outflows so that some fraction of gas becomes unbound, the total radiated energy would not be as large as expected simply based on the total mass of the disrupted stars, indicating the duration of the burst would be shortened.

For the cases with a full disruption followed by more than one partial disruption event, light curves would reveal a higher peak followed by at least one other less intense burst. If partial disruptions occur successively, the less bright bursts would show a quasi-periodic behavior with a modulation time scale $\simeq$ the orbital period of the remnants. In our simulations, the shortest period for multiple partial disruption events is $\simeq$days. In fact, in two of our models (Models with $^{\star\star}$ in Table~\ref{tab:outcome}, $M20a6\beta1/2i120$ and $M20a6\beta1/2i180$), three-body encounters create a binary system whose eccentricity is low ($e=0.74$ and $0.48$ respectively) for ordinary disruption events where the orbit is typically approximated to be parabolic. We will discuss these low-$e$ binary systems in \S~\ref{subsec:X-raybinary}.

\subsubsection{Black hole high- and low-mass X-ray binary}\label{subsec:X-raybinary}

Three-body interactions may result in the exchange of a member of the binary with the intruder object \citep{Valtonen+2006} during chaotic interactions and via the micro-Hills mechanism. The member exchange also occurs between the binary star and the BH in our simulations, resulting in a star bound to the BH (e.g., Models $M2a6\beta2i30$ and $M2a6\beta1/2i150$). In some of the models the orbit is nearly parabolic (e.g., Models $M2a2\beta1/2i30$ and $M20a6\beta1/2i30$) so that the bound star would be fully disrupted upon return. This would look exactly like an ordinary TDE. On the other hand, we also found that eccentric star-BH binaries form. In the two Models (Models $M20a6\beta1/2i120$ and $M20a6\beta1/2i180$), we actually simulate multiple quasi-periodic episodes of mass transfer at pericenter in the binary with a $\gtrsim8\Msol$ star. In Figure~\ref{fig:x_ray_binary}, we show the density distribution, projected on the $x-y$ plane, in Model $a20a6\beta1/2i120$ at a few different times until an eccentric binary forms ($a\simeq 27\Rsol$, $e=0.74$, $P\simeq 3$ days). The repeated mass loss at pericenter leads to quasi-periodic bursts in the mass accretion rate with peak $\dot{M}\simeq 10^{-5}\Msol/\sec$ on a timescale of the orbital period, as shown in Figure~\ref{fig:mdot_Xraybinary}. In that panel, we also show the accretion rate for Model $M20a6\beta1/2i180$ where an eccentric binary with $a\simeq24\Rsol$, $e\simeq 0.5$ and $P\simeq2.5$ days forms. Interestingly, this less eccentric binary does not reveal modulations of the accretion rate. Like the binary in the Model $a20a6\beta1/2i120$, the star is periodically disrupted. But because of not sufficiently large eccentricity and semimajor axis, the mass stripped from the star is simply added to an existing accretion flow around the BH. Our simulations confirm the formation of eccentric high-mass X-ray binaries in three-body encounters. 

{As shown in Figure~\ref{fig:orbit_Xraybinary}, the interacting binaries circularize; both the semimajor axis and eccentricity decrease over time. However, the two circularizing binaries show somewhat different trends in the evolution of $a$ and $e$. For the wider and more eccentric binary (blue), we see a sudden drop in both parameters (like a step function) whenever the binary undergoes a mass loss episode near pericenter. Furthermore, the absolute time derivatives of $a$ and $e$ increase slightly until the end of the simulation, indicating the circularization accelerates. On the other hand, both parameters for the smaller and more circular binaries (green) decrease like a damped oscillator at a rate that decreases over time. This may indicate that the orbit evolution of an interacting binary depends on the orbital parameters at the time two companions in the binary start to interact. } 

We also found eccentric hard binaries consisting of a $\simeq 1\Msol$ star and the 20$\Msol$ BH (Models $M2a6\beta2i30$, $M2a6\beta1i30$, $M2a6\nu45$ and $M2a6\nu135$, marked with the superscript $\star$ in Table~\ref{tab:outcome}). The orbital period of three of the binaries are 8 - 20 days and the widest one has a period of $\simeq100$ days. Two of those binaries have pericenter distances ($r_{\rm p}\simeq 3-4\Rsol$) small enough for close interactions between the star and the BH at pericenter (Models $M2a6\beta1i30$ and $M2a6\nu45$). Even the two other larger binaries are found to be hard (the orbital velocity at apocenter $\gtrsim 100\km/\sec$, which is larger than typical velocity dispersion $\simeq 10 - 15\km/\sec$ of globular clusters \citealt{Cohen+1983}), meaning that unless a strong encounter significantly disrupt the binary orbit, they may be hardened further by weak encounters to compact binaries where the star transfers its mass to the BH. 

The formation of hard binaries with a $\simeq 1\Msol$ star and a 20$\Msol$ BH may have interesting implications for the formation of detached BH-star binaries \citep{Giesers+2018} and BH-LMXBs \citep{Casares+2014}. 
Our simulations confirm the possibility that even a single strong dynamical encounter between a BH and a circular solar-type binary star can result in the formation of a BH-star binary which is already compact enough for mass transfer or is sufficiently hard that it would  potentially evolve into an X-ray binary. Considering that we have not explored the entire range of initial conditions for 3-body encounters, it will of great interest to extend
our simulations to fully
identify the parameter space where BH low-mass X-ray binaries can form. We reserve this exploration for future work.

\subsubsection{Runaway stars and isolated wandering black holes}

Runaway stars are a population of fast-moving O and B type stars at $\gtrsim 30\km/\sec$ \citep{Blaauw1961,Stone1979}. Two competing mechanisms for their formation have been suggested: (a) ejection of a star from a binary system when its companion goes off as a supernova explosion \citep[e.g.,][]{Zwicky1957, Blaauw1961}; (b) ejection of a star during dynamical interactions of binaries with other stars in a star cluster \citep[e.g.,][]{Poveda+1967,GiesBolton1986,Ryu+2017} or resulting from the interaction between infalling star clusters and massive black holes in Galactic Centres \citep[e.g.,][]{Capuzzo-DolcettaFragione2015,FragioneCapuzzo-Dolcetta2017}. 

We find that the three-body encounters involving the relatively tight binary tend to create a rapidly moving ejected star at $v\simeq 50 - 310\km/\sec$. Those velocities are large enough to escape globular clusters (the escape velocity of globular clusters $\simeq$ a few to 180 km/sec; \citealt{Antonini+2016,Gnedin+2002}).

Interestingly, in the two cases with the initially $20\Msol$ binary where one star is fully disrupted and the other star is ejected at a high speed (Models $M20a6\beta1/2i30$ and $M20a6\beta1/2i60$), the single BH gains a large momentum, moving at a high speed $\sim 60-80$~km/s. These velocities tend to be a factor of 2 - 3 smaller than the ejection velocity of the unbound star. We attribute this smaller BH velocity to the (not perfect) cancellation of the momentum kicks due to a disruption event and the ejection of a star via the slingshot mechanism. We also estimate that the angle between the unbound star and the BH is around $140-180^{\circ}$. Despite the small sample number, this suggests that the formation of a runaway star via three-body interactions between O-type stars and stellar-mass BHs may be accompanied by the formation of an active BH moving away from the unbound star at a still high speed. This finding has two main implications. First, a detection of a runaway star and an active BH moving away from each other, if the potential formation site can be identified via, e.g., integration of the trajectories backward in time, may serve as strong evidence of close three-body encounters between stars and BHs. Second, if a rapidly moving active single BH is observed, it can be used as a guide to locate a runaway star. Lastly, identifying either transient phenomena or wandering isolated BHs created due to three-body encounters can be used to mutually constrain each other, using the rates of each outcome measured by three-body scattering experiments which carefully take into account hydrodynamic effects.

\begin{figure}
	\centering
	\includegraphics[width=8.5cm,height = 6.2cm]{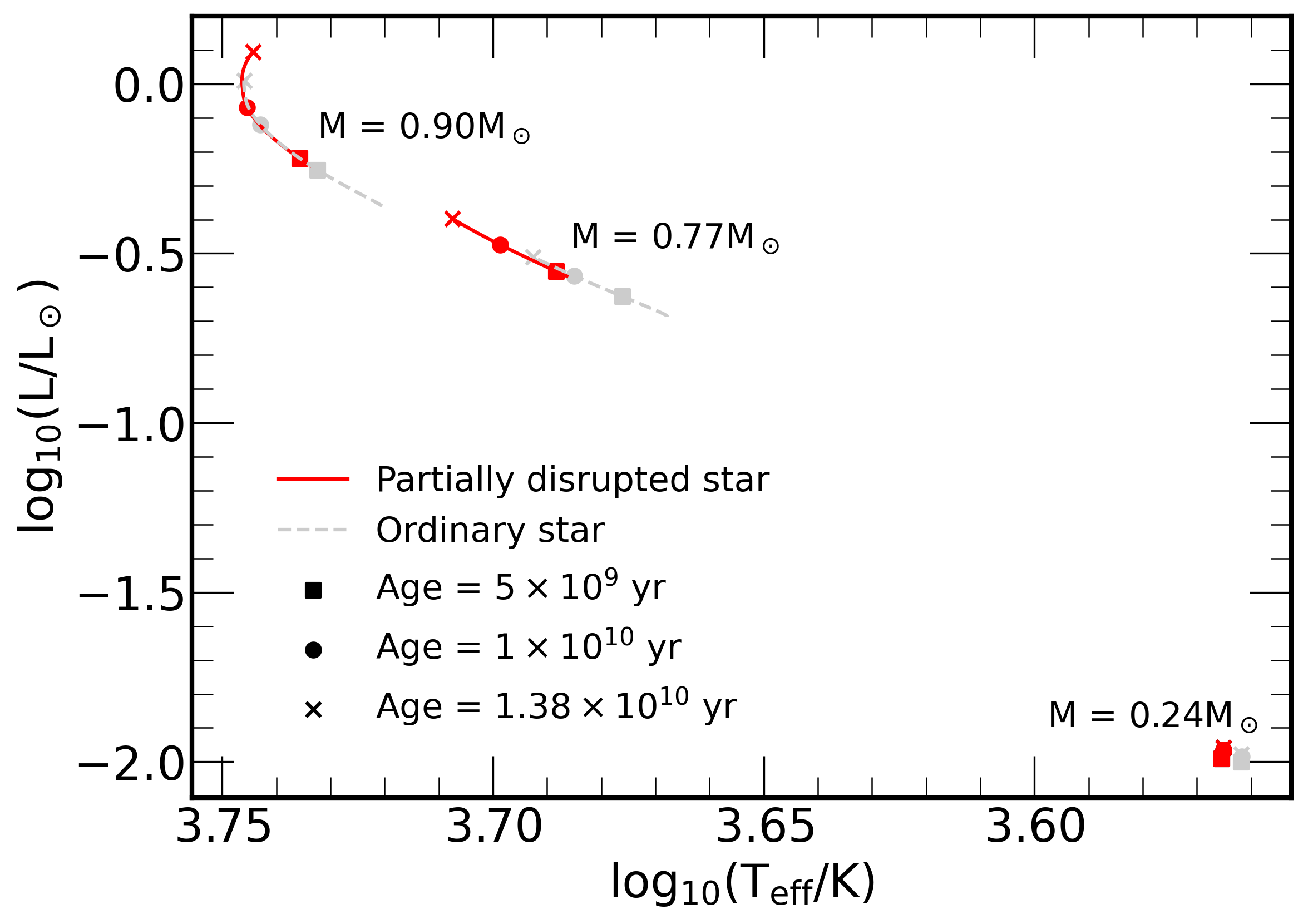}
\caption{The Hertzsprung–Russell diagram for three partially disrupted stars (red solid lines) which lost $\simeq 10\%$ ($M=0.9\Msol$, Model $M2a9\beta1/2i30$), $\simeq25\%$ ($M=0.77\Msol$, Model $M2a6\beta1/2i30$) and $\simeq 76\%$ ($M=0.24\Msol$, Model $M2a6\beta1/2i0$) of their mass, respectively. For comparison, we depict the lines for ordinary Solar-metallicity stars of the same mass using grey dotted lines. The three markers indicate the age of the star: square ($5$~Gyr), circle ($10$~Gyr) and cross (13.8~Gyr).  }
	\label{fig:HRdiagram}
\end{figure}

\subsubsection{Long-term evolution of partially disrupted stars}

Some of the runaway stars were partially disrupted before being ejected. Upon ejection, not long after the disruption event, the partially disrupted runaway stars are characterized by differential rotation and a hotter thermodynamic state. \citet{Ryu+2020c} showed similar features of partial disruption remnants by supermassive BHs. Furthermore, we find that the differential rotation and non-spherical distribution of chemical components during closest passage results in chemical mixing inside the remnants (e.g., an increase in the core H fraction by $23\%$ in a remnant of the initially $1\Msol$ star that lost 23\% of its mass).

Although the mass loss of the ejected stars during interactions with the BH is not found to be significant for the encounter parameters considered in this study, severe partial disruptions of a star, followed by its ejection, are in principle possible. Hence, given the possibility of forming runaway stars that underwent a partial disruption, we examined the long-term evolution of three partially disrupted stars: an unbound star with mass loss of $10\%$, and two bound stars with mass loss of $25\%$ and $76\%$.  As the first attempt for this approach, we only examine the impact of the additional chemical mixing on the structure of the star in this paper. We consider a non-rotating ordinary star with the post-encounter mass in \mesa~whose chemical composition profile is given by that of a partially disrupted star. We first relax the star until it reaches thermal equilibrium, and evolve it by solving the fully coupled structure and composition equations. 

Figure~\ref{fig:HRdiagram} shows the evolutionary paths of the three partially disrupted stars and ordinary Solar-metallicity stars of the same mass in the Hertzsprung-Russell diagram. The general trend is that the partially disrupted stars, once settled, are hotter and brighter than ordinary stars of the same mass at the same age (at most by a factor of a few) because of a higher He fraction. We should note that we only take the modified chemical composition profile of the partially disrupted stars into account. It is important to include all other thermodynamics quantities properly altered by tidal interactions, such as density, internal energy and angular velocity. Due to the encounter the partially disrupted stars can be spun up beyond their critical rotation rates. At this point, we would then expect the partially disrupted stars to shed angular momentum via mass loss to bring its rotation rate below critical. At this time we have ignored the effect of rotation and rotation-induced mass loss in Figure~\ref{fig:HRdiagram}. We will perform a more systematic study for this evolution with proper modelling of the internal structure, rotation, and mass-loss in our future work. 

\subsection{Event rate}\label{subsec:rate}

As an order of magnitude estimate, the differential rate of a single BH encountering a binary star may be estimated as $d\mathcal{R}/d N_{\bullet}\simeq n\Sigma v_{\rm rel}$. Here, $n$ is the the binary number density in the vicinity of the BH, $\simeq f_{\rm b}n_{\rm s}$ where $f_{\rm b}$ is the binary fraction, $\simeq 0.1$ for globular clusters  \citep{Ji2013, Ivanova+2005,Dalessandro+2011}, and $n_{\rm s}$ is the number density of single stars near the center. And $v_{\rm rel}$ is the relative velocity between the binary and the BH and $\Sigma$ is the encounter cross-section. In the gravitational focusing regime ($\sqrt{GM_{\bullet}/r_{\rm p}}\gg v_{\rm rel}\simeq \sigma$ where  $r_{\rm p}$ is closest distance between the binary and the BH and $\sigma$ is the velocity dispersion), $\Sigma \simeq  \uppi G \Mbh r_{\rm p}/\sigma^{2}$. Motivated by the fact that encounters tend to become disruptive when $r_{\rm p}< a(1+e)$ and $a\lesssim 9 a_{\rm RL}\simeq 24\Rsol$ (for $M_{\star}=1\Msol$), we consider $\Sigma \simeq \uppi G \Mbh a (1+e)/\sigma^{2}$. Then, we find that $d\mathcal{R}/dN_{\bullet}$ is expressed as
\begin{align}\label{eq:rate}
   \frac{d\mathcal{R}}{d N_{\bullet}} & \simeq \frac{n  \uppi G \Mbh a}{\sigma},\nonumber\\
              &\simeq 10^{-10} \yr^{-1} \left(\frac{f_{\rm b}}{0.1}\right) \left(\frac{n_{\rm s}}{10^{5}{\rm pc}^{-3}}\right)\left(\frac{M_{\bullet}}{40\Msol}\right)\nonumber\\
              &\times\left(\frac{a}{24\Rsol}\right) \left(\frac{\sigma}{15\km\sec}\right)^{-1}.
\end{align}
Assuming more than tens of single stellar-mass black holes existing in dense clusters at present day \citep{Morscher+2015,Askar+2018,Kremer+2020}\footnote{\citet{Askar+2018} finds that $\simeq10\%$ of the retained BH population forms a binary, which is not large enough to significantly affect our order of magnitude estimate for $\mathcal{R}$.} and $\simeq$150 globular clusters in Milky Way \citep{Harris+2010}, the rate of disruptive three-body encounters per Milky Way-like galaxy is\footnote{Note that calculating $\mathcal{R}$ involves an integration of the differential formation rate with several important factors, such star formation history. Thus for a more precise estimate of $\mathcal{R}$, a more careful consideration of cluster evolution history is required.
}, 
\begin{align}\label{eq:rate1}
   \mathcal{R}& \simeq 10^{-6} \yr^{-1} \left(\frac{N_{\bullet}}{15000}\right)\left(\frac{f_{\rm b}}{0.1}\right) \left(\frac{n_{\rm s}}{10^{5}{\rm pc}^{-3}}\right)\left(\frac{M_{\bullet}}{40\Msol}\right)\nonumber\\
              &\times\left(\frac{a}{24\Rsol}\right) \left(\frac{\sigma}{15\km\sec}\right)^{-1}.
\end{align}

We now compare the relative frequency of the three-body encounters for different binary mass $M_{\rm b}$. The number of encounters of an equal-mass binary of mass $2 M_{1}$ relative to that of an equal-mass binary of mass $2 M_{2}$ in the lifetime of the binary $t_{\rm life}$ can be expressed as $n_{1}\Sigma_{1} v_{\rm rel,1}t_{\rm life,1}/[n_{2}\Sigma_{2} v_{\rm rel,2}t_{\rm life,2}]$.

To start with, let us ignore the impact of mass segregation on the number density of stars with mass. This allows us to assume that $v_{\rm rel,1}=v_{\rm rel,2}$. Adopting the Kroupa stellar mass function \citep{Kroupa2002} with the cut-off mass $0.5\Msol$ yields $n_{1}/n_{2}=(m_{1}/m_{2})^{-2.3}(f_{\rm b, 1}/f_{\rm b,2})$. The encounter cross-section is again $\Sigma\propto r_{\rm p}$, giving  $\Sigma_{1}/\Sigma_{2}\simeq r_{\rm p,1}/r_{\rm p,2}\simeq a_{1}/a_{2}$ where $a$ is the binary semimajor axis. Finally, let us assume that $t_{\rm life}$ is comparable to the lifetime of the star $\propto m^{-3}$.

Combining all of the above we obtain
\begin{align}
    \frac{N_{1}}{N_{2}}&\simeq\frac{n_{1}\Sigma_{1} v_{\rm rel,1}t_{\rm life,1}}{n_{2}\Sigma_{2} v_{\rm rel,2}t_{\rm life,2}}\nonumber\\
            & \simeq \left(\frac{m_{1}}{m_{2}}\right)^{-2.3}\left(\frac{f_{1}}{f_{2}}\right)\left(\frac{a_{1}}{a_{2}}\right)\left(\frac{m_{1}}{m_{2}}\right)^{-3}\nonumber\\
            &\simeq \left(\frac{m_{1}}{m_{2}}\right)^{-5.3}\left(\frac{f_{1}}{f_{2}}\right)\left(\frac{a_{1}}{a_{2}}\right).
\end{align}
This suggests that the encounters of a $2\Msol$ binary (say, $a=0.05\AU$ and $f_{1}=0.1$) with a $20\Msol$ black hole are much more frequent than those of $20\Msol$ binaries (say, $a=0.1\AU$ and $f_{2}=1$).

If mass segregation is taken into account, the number density of more massive binaries around the BHs (which get to the center first) would be larger. However, $v_{\rm rel}$ would be smaller for more massive stars by the inverse square-root of the mass ratio. Also, $t_{\rm life}$ would be shortened by the formation time scale of the BH because the encounters considered only occur after BHs form. This means that $t_{\rm life}$ for more massive stars would be shortened even more. For example, the lifetime of a $\simeq 70\Msol$ star, which would collapse to a $20\Msol$ BH (Figure 2 in \citealt{SperaMapelli2017}), is $\simeq 3$ Myr. Hence $t_{\rm life}$ for a $20\Msol$ binary is 20 - 3 = 17~Myr, whereas that for a $2\Msol$ binary is $10^{4}-3\simeq 10^{4}$ Myr. This suggests that unless the number density for $20\Msol$ binaries is a few orders of magnitude larger than that for $2\Msol$ binaries, the encounters between a $2\Msol$ binary and a $20\Msol$ BH are more frequent than those between a $20\Msol$ binary and a BH of the same mass.

However, a close encounter of a BH with more massive binaries (which results in at least one full disruption) is preferentially more likely to be detected since the event is expected be brighter. The timescale of the peak luminosity will also influence the observability. However, if the timescale is proportional to the peak mass return time, then it has a weak dependence on the stellar mass; only a factor of 2 difference between a $1\Msol$ star and a $10\Msol$ star. If the luminosity has the same dependence on the stellar mass as the mass return rate, the luminosity is $\propto m^{0.7}$. Considering the strong dependence of $N_{1}/N_{2}$ on the binary mass, even the number of observable encounters is still likely to be more frequent for encounters involving low-mass stars.

\section{Summary and Conclusions}\label{sec:summary}
In this work, we investigated the outcomes of three-body encounters between a binary star and a BH using the moving-mesh hydrodynamics code {\small AREPO}. In particular, we focused on identifying all possible types of transient phenomena produced in the three-body encounters, and studied their properties. We consider a wide range of encounter parameters, i.e., the binary mass, the binary size, the impact parameter, the inclination angle, and the phase angle. However, given the inhomogeneous parameter sampling and the limited number of simulations, our work is not suitable for statistical analysis. Nonetheless, we found some clear qualitative dependence of outcomes and their properties on the system parameters. Our results can be summarized as follows. 

$\bullet$ We identified two different types of outcomes: 
\begin{enumerate}
\item \textit{Non-disruptive} encounters where neither star is disrupted, which include a weak perturbation of the binary orbit (without being dissociated) and a dissociation of the binary, creating one unbound and one bound stars. \item\textit{Disruptive} encounters where at least one of the stars is fully or partially disrupted. Prompt electromagnetic transient phenomena would be produced in the second class. Although electromagnetic radiation is not expected during encounters of the first type, the formation of hard binaries in that class indicates the possibility of electromagnetic transient phenomena from X-ray binaries at later times if the binaries' orbit continues to shrink via weak encounters. 
\end{enumerate}

$\bullet$ The most important factors to determine the parameter space for transient phenomena are the impact parameter, the binary size and the phase angle. Our simulations suggest that transients are more likely to be created in encounters of smaller binaries with smaller impact parameters. The dependence on the three parameters is:
\begin{enumerate}
    \item \textit{Impact parameter}: a large impact parameter is a necessary condition for \textit{non-disruptive} encounters whereas a small impact parameter is a sufficient condition for disruptive encounters. 
    \item \textit{Binary size}: encounters involving an initially smaller binary more likely lead to transient phenomena than non-transient phenomena because the interactions become more chaotic.  Disruption events can still happen for a wide binary if the impact parameter is $\simeq a/2$ or interactions become chaotic. 
    \item \textit{Phase angle}: the outcomes can vary, depending on the phase angle even when all other parameters are fixed.  The statistical likelihood of transient formation would be mostly governed by the binary size and impact parameter, but the phase angle is an important factor to determine the outcomes of an individual encounter case.
\end{enumerate}

$\bullet$ The transient formation in the three-body encounters has important implications:

\begin{enumerate}
    \item \textit{Varieties of transients}: three body-encounters can produce transient phenomena with unique observational features. In particular, multiple disruption events, such as instantaneous double disruption events or a full disruption event followed by repeated partial disruption events can never happen in ordinary tidal disruption events. Furthermore, a nearly collision-like disruption during chaotic interactions would have different light curves than those of ordinary tidal disruption events. Furthermore, the total energy budget available for radiation would be larger. 

    \item \textit{Black hole X-ray binary}: we found eccentric hard binaries with $\simeq1\Msol$ star form via the three-body encounters, two of which have a pericenter distance small enough for the star and BH to interact at pericenter. The formation of these particular star-BH binary systems may serve as evidence for the dynamical formation of BH low-mass X-ray binaries whose formation mechanism remains questionable. On the other hand, we found that a compact nearly equal-mass star-BH binary with mass $40\Msol$ can form, confirming the formation of high-mass X-ray binaries via three-body interactions between a binary star and a BH.

    \item \textit{Runaway stars and isolated black holes}: one frequent outcome in our simulations is the formation of single (both $1\Msol$ and $10\Msol$) stars ejected at $50-270\km/\sec$ via chaotic interactions or the Hills mechanism \citep{Hills1988} (which we call 'micro-Hills' mechanism). In particular, the velocities of the ejected stars with $M_{\star}\simeq 10\Msol$ are comparable to typical velocities of runaway stars ($\gtrsim 30\km/\sec$). In the two cases where one star is ejected at a high speed and the other star is fully disrupted, the single BH is also ejected at a velocity of $30-80\km/\sec$ in the opposite direction to the unbound star, becoming active due to the accretion of the surrounding stellar debris. A detection of a runaway star or wandering active BH could mutually constrain the population of each other. 
\end{enumerate}

In \citetalias{Ryu+2022}, we investigated the production of transient events in three-body encounters between a single star and a binary BH using hydrodynamics simulations. We found that various types of transients can form in each particular type of encounter, with their own unique observational signatures, such as light curves with periodic modulations. This work further adds to the significance of three-body encounters as a formation channel of transient phenomena.

\section*{Acknowledgements}
The authors are grateful to the anonymous referee for constructive suggestions and comments which helped us improve the manuscript. This research project was conducted using computational resources (and/or scientific computing services) at the Max-Planck Computing \& Data Facility. The authors would like to also thank Stony Brook Research Computing and Cyberinfrastructure, and the Institute for Advanced Computational Science at Stony Brook University for access to the high-performance SeaWulf computing system, which was made possible by a \$1.4M National Science Foundation grant (\#1531492).
R.Perna acknowledges support by NSF award AST-2006839.

\section*{Data Availability}
Any data used in this analysis are available on reasonable request from the first author.

\bibliographystyle{mnras}
%\bibliography{biblio.bib} 

\begin{thebibliography}{}
	\makeatletter
	\relax
	\def\mn@urlcharsother{\let\do\@makeother \do\$\do\&\do\#\do\^\do\_\do\%\do\~}
	\def\mn@doi{\begingroup\mn@urlcharsother \@ifnextchar [ {\mn@doi@}
		{\mn@doi@[]}}
	\def\mn@doi@[#1]#2{\def\@tempa{#1}\ifx\@tempa\@empty \href
		{http://dx.doi.org/#2} {doi:#2}\else \href {http://dx.doi.org/#2} {#1}\fi
		\endgroup}
	\def\mn@eprint#1#2{\mn@eprint@#1:#2::\@nil}
	\def\mn@eprint@arXiv#1{\href {http://arxiv.org/abs/#1} {{\tt arXiv:#1}}}
	\def\mn@eprint@dblp#1{\href {http://dblp.uni-trier.de/rec/bibtex/#1.xml}
		{dblp:#1}}
	\def\mn@eprint@#1:#2:#3:#4\@nil{\def\@tempa {#1}\def\@tempb {#2}\def\@tempc
		{#3}\ifx \@tempc \@empty \let \@tempc \@tempb \let \@tempb \@tempa \fi \ifx
		\@tempb \@empty \def\@tempb {arXiv}\fi \@ifundefined
		{mn@eprint@\@tempb}{\@tempb:\@tempc}{\expandafter \expandafter \csname
			mn@eprint@\@tempb\endcsname \expandafter{\@tempc}}}
	
	\bibitem[\protect\citeauthoryear{{Antonini} \& {Rasio}}{{Antonini} \&
		{Rasio}}{2016}]{Antonini+2016}
	{Antonini} F.,  {Rasio} F.~A.,  2016, \mn@doi [\apj]
	{10.3847/0004-637X/831/2/187}, \href
	{https://ui.adsabs.harvard.edu/abs/2016ApJ...831..187A} {831, 187}
	
	\bibitem[\protect\citeauthoryear{{Askar}, {Arca Sedda}  \& {Giersz}}{{Askar}
		et~al.}{2018}]{Askar+2018}
	{Askar} A.,  {Arca Sedda} M.,   {Giersz} M.,  2018, \mn@doi [\mnras]
	{10.1093/mnras/sty1186}, \href
	{https://ui.adsabs.harvard.edu/abs/2018MNRAS.478.1844A} {478, 1844}
	
	\bibitem[\protect\citeauthoryear{{Blaauw}}{{Blaauw}}{1961}]{Blaauw1961}
	{Blaauw} A.,  1961, \bain, \href
	{https://ui.adsabs.harvard.edu/abs/1961BAN....15..265B} {15, 265}
	
	\bibitem[\protect\citeauthoryear{{Capuzzo-Dolcetta} \&
		{Fragione}}{{Capuzzo-Dolcetta} \&
		{Fragione}}{2015}]{Capuzzo-DolcettaFragione2015}
	{Capuzzo-Dolcetta} R.,  {Fragione} G.,  2015, \mn@doi [\mnras]
	{10.1093/mnras/stv2123}, \href
	{https://ui.adsabs.harvard.edu/abs/2015MNRAS.454.2677C} {454, 2677}
	
	\bibitem[\protect\citeauthoryear{{Casares} \& {Jonker}}{{Casares} \&
		{Jonker}}{2014}]{Casares+2014}
	{Casares} J.,  {Jonker} P.~G.,  2014, \mn@doi [\ssr]
	{10.1007/s11214-013-0030-6}, \href
	{https://ui.adsabs.harvard.edu/abs/2014SSRv..183..223C} {183, 223}
	
	\bibitem[\protect\citeauthoryear{{Chomiuk}, {Strader}, {Maccarone},
		{Miller-Jones}, {Heinke}, {Noyola}, {Seth}  \& {Ransom}}{{Chomiuk}
		et~al.}{2013}]{Chomiuk+2013}
	{Chomiuk} L.,  {Strader} J.,  {Maccarone} T.~J.,  {Miller-Jones} J. C.~A.,
	{Heinke} C.,  {Noyola} E.,  {Seth} A.~C.,   {Ransom} S.,  2013, \mn@doi
	[\apj] {10.1088/0004-637X/777/1/69}, \href
	{https://ui.adsabs.harvard.edu/abs/2013ApJ...777...69C} {777, 69}
	
	\bibitem[\protect\citeauthoryear{{Cohen}}{{Cohen}}{1983}]{Cohen+1983}
	{Cohen} J.~G.,  1983, \mn@doi [\apjl] {10.1086/184066}, \href
	{https://ui.adsabs.harvard.edu/abs/1983ApJ...270L..41C} {270, L41}
	
	\bibitem[\protect\citeauthoryear{{Corral-Santana}, {Casares},
		{Mu{\~n}oz-Darias}, {Bauer}, {Mart{\'\i}nez-Pais}  \&
		{Russell}}{{Corral-Santana} et~al.}{2016}]{Corral-Santana+2016}
	{Corral-Santana} J.~M.,  {Casares} J.,  {Mu{\~n}oz-Darias} T.,  {Bauer} F.~E.,
	{Mart{\'\i}nez-Pais} I.~G.,   {Russell} D.~M.,  2016, \mn@doi [\aap]
	{10.1051/0004-6361/201527130}, \href
	{https://ui.adsabs.harvard.edu/abs/2016A&A...587A..61C} {587, A61}
	
	\bibitem[\protect\citeauthoryear{{Dalessandro}, {Lanzoni}, {Beccari},
		{Sollima}, {Ferraro}  \& {Pasquato}}{{Dalessandro}
		et~al.}{2011}]{Dalessandro+2011}
	{Dalessandro} E.,  {Lanzoni} B.,  {Beccari} G.,  {Sollima} A.,  {Ferraro}
	F.~R.,   {Pasquato} M.,  2011, \mn@doi [\apj] {10.1088/0004-637X/743/1/11},
	\href {https://ui.adsabs.harvard.edu/abs/2011ApJ...743...11D} {743, 11}
	
	\bibitem[\protect\citeauthoryear{{Duch{\^e}ne} \& {Kraus}}{{Duch{\^e}ne} \&
		{Kraus}}{2013}]{DucheneKraus2013}
	{Duch{\^e}ne} G.,  {Kraus} A.,  2013, \mn@doi [\araa]
	{10.1146/annurev-astro-081710-102602}, \href
	{https://ui.adsabs.harvard.edu/abs/2013ARA&A..51..269D} {51, 269}
	
	\bibitem[\protect\citeauthoryear{{Eggleton}}{{Eggleton}}{1983}]{Eggleton1983}
	{Eggleton} P.~P.,  1983, \mn@doi [\apj] {10.1086/160960}, \href
	{https://ui.adsabs.harvard.edu/abs/1983ApJ...268..368E} {268, 368}
	
	\bibitem[\protect\citeauthoryear{{Fanidakis}, {Baugh}, {Benson}, {Bower},
		{Cole}, {Done}  \& {Frenk}}{{Fanidakis} et~al.}{2011}]{Fanidakis+2011}
	{Fanidakis} N.,  {Baugh} C.~M.,  {Benson} A.~J.,  {Bower} R.~G.,  {Cole} S.,
	{Done} C.,   {Frenk} C.~S.,  2011, \mn@doi [\mnras]
	{10.1111/j.1365-2966.2010.17427.x}, \href
	{https://ui.adsabs.harvard.edu/abs/2011MNRAS.410...53F} {410, 53}
	
	\bibitem[\protect\citeauthoryear{{Fragione}, {Capuzzo-Dolcetta}  \&
		{Kroupa}}{{Fragione} et~al.}{2017}]{FragioneCapuzzo-Dolcetta2017}
	{Fragione} G.,  {Capuzzo-Dolcetta} R.,   {Kroupa} P.,  2017, \mn@doi [\mnras]
	{10.1093/mnras/stx106}, \href
	{https://ui.adsabs.harvard.edu/abs/2017MNRAS.467..451F} {467, 451}
	
	\bibitem[\protect\citeauthoryear{{Fragione}, {Perna}  \& {Loeb}}{{Fragione}
		et~al.}{2021}]{Fragione2021}
	{Fragione} G.,  {Perna} R.,   {Loeb} A.,  2021, \mn@doi [\mnras]
	{10.1093/mnras/staa3493}, \href
	{https://ui.adsabs.harvard.edu/abs/2021MNRAS.500.4307F} {500, 4307}
	
	\bibitem[\protect\citeauthoryear{{Gies} \& {Bolton}}{{Gies} \&
		{Bolton}}{1986}]{GiesBolton1986}
	{Gies} D.~R.,  {Bolton} C.~T.,  1986, \mn@doi [\apjs] {10.1086/191118}, \href
	{https://ui.adsabs.harvard.edu/abs/1986ApJS...61..419G} {61, 419}
	
	\bibitem[\protect\citeauthoryear{{Giesers} et~al.,}{{Giesers}
		et~al.}{2018}]{Giesers+2018}
	{Giesers} B.,  et~al., 2018, \mn@doi [\mnras] {10.1093/mnrasl/slx203}, \href
	{https://ui.adsabs.harvard.edu/abs/2018MNRAS.475L..15G} {475, L15}
	
	\bibitem[\protect\citeauthoryear{{Gnedin}, {Zhao}, {Pringle}, {Fall}, {Livio}
		\& {Meylan}}{{Gnedin} et~al.}{2002}]{Gnedin+2002}
	{Gnedin} O.~Y.,  {Zhao} H.,  {Pringle} J.~E.,  {Fall} S.~M.,  {Livio} M.,
	{Meylan} G.,  2002, \mn@doi [\apjl] {10.1086/340319}, \href
	{https://ui.adsabs.harvard.edu/abs/2002ApJ...568L..23G} {568, L23}
	
	\bibitem[\protect\citeauthoryear{{Goodman} \& {Hernquist}}{{Goodman} \&
		{Hernquist}}{1991}]{GoodmanHernquist1991}
	{Goodman} J.,  {Hernquist} L.,  1991, \mn@doi [\apj] {10.1086/170464}, \href
	{https://ui.adsabs.harvard.edu/abs/1991ApJ...378..637G} {378, 637}
	
	\bibitem[\protect\citeauthoryear{{Harris}}{{Harris}}{2010}]{Harris+2010}
	{Harris} W.~E.,  2010, arXiv e-prints, \href
	{https://ui.adsabs.harvard.edu/abs/2010arXiv1012.3224H} {p. arXiv:1012.3224}
	
	\bibitem[\protect\citeauthoryear{{Hills}}{{Hills}}{1988}]{Hills1988}
	{Hills} J.~G.,  1988, \mn@doi [\nat] {10.1038/331687a0}, \href
	{https://ui.adsabs.harvard.edu/abs/1988Natur.331..687H} {331, 687}
	
	\bibitem[\protect\citeauthoryear{{Hut} et~al.,}{{Hut} et~al.}{1992}]{Hut1992}
	{Hut} P.,  et~al., 1992, \mn@doi [\pasp] {10.1086/133085}, \href
	{https://ui.adsabs.harvard.edu/abs/1992PASP..104..981H} {104, 981}
	
	\bibitem[\protect\citeauthoryear{{Ivanova}, {Belczynski}, {Fregeau}  \&
		{Rasio}}{{Ivanova} et~al.}{2005}]{Ivanova+2005}
	{Ivanova} N.,  {Belczynski} K.,  {Fregeau} J.~M.,   {Rasio} F.~A.,  2005,
	\mn@doi [\mnras] {10.1111/j.1365-2966.2005.08804.x}, \href
	{https://ui.adsabs.harvard.edu/abs/2005MNRAS.358..572I} {358, 572}
	
	\bibitem[\protect\citeauthoryear{{Jermyn} et~al.,}{{Jermyn}
		et~al.}{2022}]{jermyn22}
	{Jermyn} A.~S.,  et~al., 2022, arXiv e-prints, \href
	{https://ui.adsabs.harvard.edu/abs/2022arXiv220803651J} {p. arXiv:2208.03651}
	
	\bibitem[\protect\citeauthoryear{{Ji} \& {Bregman}}{{Ji} \&
		{Bregman}}{2013}]{Ji2013}
	{Ji} J.,  {Bregman} J.~N.,  2013, \mn@doi [\apj] {10.1088/0004-637X/768/2/158},
	\href {https://ui.adsabs.harvard.edu/abs/2013ApJ...768..158J} {768, 158}
	
	\bibitem[\protect\citeauthoryear{{King}, {Kolb}  \& {Burderi}}{{King}
		et~al.}{1996}]{KingBurderi1996}
	{King} A.~R.,  {Kolb} U.,   {Burderi} L.,  1996, \mn@doi [\apjl]
	{10.1086/310105}, \href
	{https://ui.adsabs.harvard.edu/abs/1996ApJ...464L.127K} {464, L127}
	
	\bibitem[\protect\citeauthoryear{{Kremer}, {Chatterjee}, {Rodriguez}  \&
		{Rasio}}{{Kremer} et~al.}{2018a}]{Kremer+2018c}
	{Kremer} K.,  {Chatterjee} S.,  {Rodriguez} C.~L.,   {Rasio} F.~A.,  2018a,
	\mn@doi [\apj] {10.3847/1538-4357/aa99df}, \href
	{https://ui.adsabs.harvard.edu/abs/2018ApJ...852...29K} {852, 29}
	
	\bibitem[\protect\citeauthoryear{{Kremer}, {Ye}, {Chatterjee}, {Rodriguez}  \&
		{Rasio}}{{Kremer} et~al.}{2018b}]{Kremer+2018}
	{Kremer} K.,  {Ye} C.~S.,  {Chatterjee} S.,  {Rodriguez} C.~L.,   {Rasio}
	F.~A.,  2018b, \mn@doi [\apjl] {10.3847/2041-8213/aab26c}, \href
	{https://ui.adsabs.harvard.edu/abs/2018ApJ...855L..15K} {855, L15}
	
	\bibitem[\protect\citeauthoryear{{Kremer}, {Lu}, {Rodriguez}, {Lachat}  \&
		{Rasio}}{{Kremer} et~al.}{2019}]{Kremer2019}
	{Kremer} K.,  {Lu} W.,  {Rodriguez} C.~L.,  {Lachat} M.,   {Rasio} F. c.~A.,
	2019, \mn@doi [\apj] {10.3847/1538-4357/ab2e0c}, \href
	{https://ui.adsabs.harvard.edu/abs/2019ApJ...881...75K} {881, 75}
	
	\bibitem[\protect\citeauthoryear{{Kremer} et~al.,}{{Kremer}
		et~al.}{2020}]{Kremer+2020}
	{Kremer} K.,  et~al., 2020, \mn@doi [\apjs] {10.3847/1538-4365/ab7919}, \href
	{https://ui.adsabs.harvard.edu/abs/2020ApJS..247...48K} {247, 48}
	
	\bibitem[\protect\citeauthoryear{{Kremer}, {Lu}, {Piro}, {Chatterjee}, {Rasio}
		\& {Ye}}{{Kremer} et~al.}{2021}]{Kremer2021}
	{Kremer} K.,  {Lu} W.,  {Piro} A.~L.,  {Chatterjee} S.,  {Rasio} F. c.~A.,
	{Ye} C.~S.,  2021, \mn@doi [\apj] {10.3847/1538-4357/abeb14}, \href
	{https://ui.adsabs.harvard.edu/abs/2021ApJ...911..104K} {911, 104}
	
	\bibitem[\protect\citeauthoryear{{Kremer}, {Lombardi}, {Lu}, {Piro}  \&
		{Rasio}}{{Kremer} et~al.}{2022}]{Kremer2022}
	{Kremer} K.,  {Lombardi} James~C. J.,  {Lu} W.,  {Piro} A.~L.,   {Rasio} F.
	e.~A.,  2022, arXiv e-prints, \href
	{https://ui.adsabs.harvard.edu/abs/2022arXiv220112368K} {p. arXiv:2201.12368}
	
	\bibitem[\protect\citeauthoryear{{Kroupa}}{{Kroupa}}{2002}]{Kroupa2002}
	{Kroupa} P.,  2002, \mn@doi [Science] {10.1126/science.1067524}, \href
	{https://ui.adsabs.harvard.edu/abs/2002Sci...295...82K} {295, 82}
	
	\bibitem[\protect\citeauthoryear{{Lopez}, {Batta}, {Ramirez-Ruiz}, {Martinez}
		\& {Samsing\ }}{{Lopez} et~al.}{2019a}]{Lopez2019}
	{Lopez} Martin J.,  {Batta} A.,  {Ramirez-Ruiz} E.,  {Martinez} I.,   {Samsing\
	} J.,  2019a, \mn@doi [\apj] {10.3847/1538-4357/ab1842}, \href
	{https://ui.adsabs.harvard.edu/abs/2019ApJ...877...56L} {877, 56}
	
	\bibitem[\protect\citeauthoryear{{Lopez}, {Batta}, {Ramirez-Ruiz}, {Martinez}
		\& {Samsing}}{{Lopez} et~al.}{2019b}]{Lopez+2019}
	{Lopez} Martin J.,  {Batta} A.,  {Ramirez-Ruiz} E.,  {Martinez} I.,   {Samsing}
	J.,  2019b, \mn@doi [\apj] {10.3847/1538-4357/ab1842}, \href
	{https://ui.adsabs.harvard.edu/abs/2019ApJ...877...56L} {877, 56}
	
	\bibitem[\protect\citeauthoryear{{McMillan}, {Cranmer}, {Shorter}  \&
		{Hernquist}}{{McMillan} et~al.}{1991}]{McMillan+1991}
	{McMillan} S. L.~W.,  {Cranmer} S.~R.,  {Shorter} S.~A.,   {Hernquist} L.,
	1991, in {Janes} K.,  ed.,  Astronomical Society of the Pacific Conference
	Series Vol. 13, The Formation and Evolution of Star Clusters. pp 418--420
	
	\bibitem[\protect\citeauthoryear{{Meibom} \& {Mathieu}}{{Meibom} \&
		{Mathieu}}{2005}]{Meibom+Mathieu2005}
	{Meibom} S.,  {Mathieu} R.~D.,  2005, \mn@doi [\apj] {10.1086/427082}, \href
	{https://ui.adsabs.harvard.edu/abs/2005ApJ...620..970M} {620, 970}
	
	\bibitem[\protect\citeauthoryear{{Michaely} \& {Perets}}{{Michaely} \&
		{Perets}}{2016}]{MichaelyPerets2016}
	{Michaely} E.,  {Perets} H.~B.,  2016, \mn@doi [\mnras] {10.1093/mnras/stw368},
	\href {https://ui.adsabs.harvard.edu/abs/2016MNRAS.458.4188M} {458, 4188}
	
	\bibitem[\protect\citeauthoryear{{Miller-Jones} et~al.,}{{Miller-Jones}
		et~al.}{2015}]{Miller-Jones+2015}
	{Miller-Jones} J.~C.~A.,  et~al., 2015, \mn@doi [\mnras]
	{10.1093/mnras/stv1869}, \href
	{https://ui.adsabs.harvard.edu/abs/2015MNRAS.453.3918M} {453, 3918}
	
	\bibitem[\protect\citeauthoryear{{Moe} \& {Di Stefano}}{{Moe} \& {Di
			Stefano}}{2017}]{Moe2017}
	{Moe} M.,  {Di Stefano} R.,  2017, \mn@doi [\apjs] {10.3847/1538-4365/aa6fb6},
	\href {https://ui.adsabs.harvard.edu/abs/2017ApJS..230...15M} {230, 15}
	
	\bibitem[\protect\citeauthoryear{{Monaghan} \& {Lattanzio}}{{Monaghan} \&
		{Lattanzio}}{1985}]{MonaghanLattanzio1985}
	{Monaghan} J.~J.,  {Lattanzio} J.~C.,  1985, \aap, \href
	{https://ui.adsabs.harvard.edu/abs/1985A&A...149..135M} {149, 135}
	
	\bibitem[\protect\citeauthoryear{{Morscher}, {Pattabiraman}, {Rodriguez},
		{Rasio}  \& {Umbreit}}{{Morscher} et~al.}{2015}]{Morscher+2015}
	{Morscher} M.,  {Pattabiraman} B.,  {Rodriguez} C.,  {Rasio} F.~A.,   {Umbreit}
	S.,  2015, \mn@doi [\apj] {10.1088/0004-637X/800/1/9}, \href
	{https://ui.adsabs.harvard.edu/abs/2015ApJ...800....9M} {800, 9}
	
	\bibitem[\protect\citeauthoryear{{Naoz}, {Fragos}, {Geller}, {Stephan}  \&
		{Rasio}}{{Naoz} et~al.}{2016}]{Naoz+2016}
	{Naoz} S.,  {Fragos} T.,  {Geller} A.,  {Stephan} A.~P.,   {Rasio} F.~A.,
	2016, \mn@doi [\apjl] {10.3847/2041-8205/822/2/L24}, \href
	{https://ui.adsabs.harvard.edu/abs/2016ApJ...822L..24N} {822, L24}
	
	\bibitem[\protect\citeauthoryear{{Ohlmann}, {R{\"o}pke}, {Pakmor}  \&
		{Springel}}{{Ohlmann} et~al.}{2017}]{Ohlmann+2017}
	{Ohlmann} S.~T.,  {R{\"o}pke} F.~K.,  {Pakmor} R.,   {Springel} V.,  2017,
	\mn@doi [\aap] {10.1051/0004-6361/201629692}, \href
	{https://ui.adsabs.harvard.edu/abs/2017A&A...599A...5O} {599, A5}
	
	\bibitem[\protect\citeauthoryear{{Pakmor}, {Edelmann}, {R{\"o}pke}  \&
		{Hillebrandt}}{{Pakmor} et~al.}{2012}]{Pakmor+2012}
	{Pakmor} R.,  {Edelmann} P.,  {R{\"o}pke} F.~K.,   {Hillebrandt} W.,  2012,
	\mn@doi [\mnras] {10.1111/j.1365-2966.2012.21383.x}, \href
	{https://ui.adsabs.harvard.edu/abs/2012MNRAS.424.2222P} {424, 2222}
	
	\bibitem[\protect\citeauthoryear{{Pakmor}, {Springel}, {Bauer}, {Mocz},
		{Munoz}, {Ohlmann}, {Schaal}  \& {Zhu}}{{Pakmor} et~al.}{2016}]{ArepoHydro}
	{Pakmor} R.,  {Springel} V.,  {Bauer} A.,  {Mocz} P.,  {Munoz} D.~J.,
	{Ohlmann} S.~T.,  {Schaal} K.,   {Zhu} C.,  2016, \mn@doi [\mnras]
	{10.1093/mnras/stv2380}, \href
	{https://ui.adsabs.harvard.edu/abs/2016MNRAS.455.1134P} {455, 1134}
	
	\bibitem[\protect\citeauthoryear{{Paxton}, {Bildsten}, {Dotter}, {Herwig},
		{Lesaffre}  \& {Timmes}}{{Paxton} et~al.}{2011}]{Paxton+2011}
	{Paxton} B.,  {Bildsten} L.,  {Dotter} A.,  {Herwig} F.,  {Lesaffre} P.,
	{Timmes} F.,  2011, \mn@doi [\apjs] {10.1088/0067-0049/192/1/3}, \href
	{http://adsabs.harvard.edu/abs/2011ApJS..192....3P} {192, 3}
	
	\bibitem[\protect\citeauthoryear{{Paxton} et~al.,}{{Paxton}
		et~al.}{2013}]{paxton:13}
	{Paxton} B.,  et~al., 2013, \mn@doi [\apjs] {10.1088/0067-0049/208/1/4}, \href
	{http://adsabs.harvard.edu/abs/2013ApJS..208....4P} {208, 4}
	
	\bibitem[\protect\citeauthoryear{{Paxton} et~al.,}{{Paxton}
		et~al.}{2015}]{paxton:15}
	{Paxton} B.,  et~al., 2015, \mn@doi [\apjs] {10.1088/0067-0049/220/1/15}, \href
	{http://adsabs.harvard.edu/abs/2015ApJS..220...15P} {220, 15}
	
	\bibitem[\protect\citeauthoryear{{Paxton} et~al.,}{{Paxton}
		et~al.}{2018}]{MESArelaxation}
	{Paxton} B.,  et~al., 2018, \mn@doi [\apjs] {10.3847/1538-4365/aaa5a8}, \href
	{https://ui.adsabs.harvard.edu/abs/2018ApJS..234...34P} {234, 34}
	
	\bibitem[\protect\citeauthoryear{{Paxton} et~al.,}{{Paxton}
		et~al.}{2019}]{paxton:19}
	{Paxton} B.,  et~al., 2019, \mn@doi [\apjs] {10.3847/1538-4365/ab2241}, \href
	{https://ui.adsabs.harvard.edu/abs/2019ApJS..243...10P} {243, 10}
	
	\bibitem[\protect\citeauthoryear{{Perets}, {Li}, {Lombardi}  \&
		{Milcarek}}{{Perets} et~al.}{2016}]{Perets2016}
	{Perets} H.~B.,  {Li} Z.,  {Lombardi} James~C. J.,   {Milcarek} Stephen~R. J.,
	2016, \mn@doi [\apj] {10.3847/0004-637X/823/2/113}, \href
	{https://ui.adsabs.harvard.edu/abs/2016ApJ...823..113P} {823, 113}
	
	\bibitem[\protect\citeauthoryear{{Podsiadlowski}, {Rappaport}  \&
		{Han}}{{Podsiadlowski} et~al.}{2003}]{Podsiadlowski+2003}
	{Podsiadlowski} P.,  {Rappaport} S.,   {Han} Z.,  2003, \mn@doi [\mnras]
	{10.1046/j.1365-8711.2003.06464.x}, \href
	{https://ui.adsabs.harvard.edu/abs/2003MNRAS.341..385P} {341, 385}
	
	\bibitem[\protect\citeauthoryear{{Poveda}, {Ruiz}  \& {Allen}}{{Poveda}
		et~al.}{1967}]{Poveda+1967}
	{Poveda} A.,  {Ruiz} J.,   {Allen} C.,  1967, Boletin de los Observatorios
	Tonantzintla y Tacubaya, \href
	{https://ui.adsabs.harvard.edu/abs/1967BOTT....4...86P} {4, 86}
	
	\bibitem[\protect\citeauthoryear{{Ryu}, {Leigh}  \& {Perna}}{{Ryu}
		et~al.}{2017}]{Ryu+2017}
	{Ryu} T.,  {Leigh} N. W.~C.,   {Perna} R.,  2017, \mn@doi [\mnras]
	{10.1093/mnras/stx1408}, \href
	{https://ui.adsabs.harvard.edu/abs/2017MNRAS.470.3049R} {470, 3049}
	
	\bibitem[\protect\citeauthoryear{{Ryu}, {Krolik}, {Piran}  \& {Noble}}{{Ryu}
		et~al.}{2020a}]{Ryu+2020a}
	{Ryu} T.,  {Krolik} J.,  {Piran} T.,   {Noble} S.~C.,  2020a, \mn@doi [\apj]
	{10.3847/1538-4357/abb3cf}, \href
	{https://ui.adsabs.harvard.edu/abs/2020ApJ...904...98R} {904, 98}
	
	\bibitem[\protect\citeauthoryear{{Ryu}, {Krolik}, {Piran}  \& {Noble}}{{Ryu}
		et~al.}{2020b}]{Ryu+2020b}
	{Ryu} T.,  {Krolik} J.,  {Piran} T.,   {Noble} S.~C.,  2020b, \mn@doi [\apj]
	{10.3847/1538-4357/abb3cd}, \href
	{https://ui.adsabs.harvard.edu/abs/2020ApJ...904...99R} {904, 99}
	
	\bibitem[\protect\citeauthoryear{{Ryu}, {Krolik}, {Piran}  \& {Noble}}{{Ryu}
		et~al.}{2020c}]{Ryu+2020c}
	{Ryu} T.,  {Krolik} J.,  {Piran} T.,   {Noble} S.~C.,  2020c, \mn@doi [\apj]
	{10.3847/1538-4357/abb3ce}, \href
	{https://ui.adsabs.harvard.edu/abs/2020ApJ...904..100R} {904, 100}
	
	\bibitem[\protect\citeauthoryear{{Ryu}, {Krolik}, {Piran}  \& {Noble}}{{Ryu}
		et~al.}{2020d}]{Ryu+2020d}
	{Ryu} T.,  {Krolik} J.,  {Piran} T.,   {Noble} S.~C.,  2020d, \mn@doi [\apj]
	{10.3847/1538-4357/abb3cc}, \href
	{https://ui.adsabs.harvard.edu/abs/2020ApJ...904..101R} {904, 101}
	
	\bibitem[\protect\citeauthoryear{{Ryu}, {Perna}  \& {Wang}}{{Ryu}
		et~al.}{2022}]{Ryu+2022}
	{Ryu} T.,  {Perna} R.,   {Wang} Y.-H.,  2022, \mn@doi [\mnras]
	{10.1093/mnras/stac2316}, \href
	{https://ui.adsabs.harvard.edu/abs/2022MNRAS.516.2204R} {516, 2204}
	
	\bibitem[\protect\citeauthoryear{{Sana} et~al.,}{{Sana}
		et~al.}{2012}]{Sana+2012}
	{Sana} H.,  et~al., 2012, \mn@doi [Science] {10.1126/science.1223344}, \href
	{https://ui.adsabs.harvard.edu/abs/2012Sci...337..444S} {337, 444}
	
	\bibitem[\protect\citeauthoryear{{Savonije}}{{Savonije}}{1978}]{Savonije+1978}
	{Savonije} G.~J.,  1978, \aap, \href
	{https://ui.adsabs.harvard.edu/abs/1978A&A....62..317S} {62, 317}
	
	\bibitem[\protect\citeauthoryear{{Secunda}, {Hernandez}, {Goodman}, {Leigh},
		{McKernan}, {Ford}  \& {Adorno}}{{Secunda} et~al.}{2021}]{Secunda2021}
	{Secunda} A.,  {Hernandez} B.,  {Goodman} J.,  {Leigh} N. W.~C.,  {McKernan}
	B.,  {Ford} K.~E.~S.,   {Adorno} J.~I.,  2021, \mn@doi [\apjl]
	{10.3847/2041-8213/abe11d}, \href
	{https://ui.adsabs.harvard.edu/abs/2021ApJ...908L..27S} {908, L27}
	
	\bibitem[\protect\citeauthoryear{{Shishkovsky} et~al.,}{{Shishkovsky}
		et~al.}{2018}]{Shishkovsky+2018}
	{Shishkovsky} L.,  et~al., 2018, \mn@doi [\apj] {10.3847/1538-4357/aaadb1},
	\href {https://ui.adsabs.harvard.edu/abs/2018ApJ...855...55S} {855, 55}
	
	\bibitem[\protect\citeauthoryear{{S{\k{a}}dowski}, {Narayan}, {McKinney}  \&
		{Tchekhovskoy}}{{S{\k{a}}dowski} et~al.}{2014}]{Skadowski2014}
	{S{\k{a}}dowski} A.,  {Narayan} R.,  {McKinney} J.~C.,   {Tchekhovskoy} A.,
	2014, \mn@doi [\mnras] {10.1093/mnras/stt2479}, \href
	{https://ui.adsabs.harvard.edu/abs/2014MNRAS.439..503S} {439, 503}
	
	\bibitem[\protect\citeauthoryear{{Sollima}, {Carballo-Bello}, {Beccari},
		{Ferraro}, {Pecci}  \& {Lanzoni}}{{Sollima} et~al.}{2010}]{Sollima+2010}
	{Sollima} A.,  {Carballo-Bello} J.~A.,  {Beccari} G.,  {Ferraro} F.~R.,
	{Pecci} F.~F.,   {Lanzoni} B.,  2010, \mn@doi [\mnras]
	{10.1111/j.1365-2966.2009.15676.x}, \href
	{https://ui.adsabs.harvard.edu/abs/2010MNRAS.401..577S} {401, 577}
	
	\bibitem[\protect\citeauthoryear{{Spera} \& {Mapelli}}{{Spera} \&
		{Mapelli}}{2017}]{SperaMapelli2017}
	{Spera} M.,  {Mapelli} M.,  2017, \mn@doi [\mnras] {10.1093/mnras/stx1576},
	\href {https://ui.adsabs.harvard.edu/abs/2017MNRAS.470.4739S} {470, 4739}
	
	\bibitem[\protect\citeauthoryear{{Springel}}{{Springel}}{2005}]{GADGET2}
	{Springel} V.,  2005, \mn@doi [\mnras] {10.1111/j.1365-2966.2005.09655.x},
	\href {https://ui.adsabs.harvard.edu/abs/2005MNRAS.364.1105S} {364, 1105}
	
	\bibitem[\protect\citeauthoryear{{Springel}}{{Springel}}{2010}]{Arepo}
	{Springel} V.,  2010, \mn@doi [\mnras] {10.1111/j.1365-2966.2009.15715.x},
	\href {https://ui.adsabs.harvard.edu/abs/2010MNRAS.401..791S} {401, 791}
	
	\bibitem[\protect\citeauthoryear{{Stone}}{{Stone}}{1979}]{Stone1979}
	{Stone} R.~C.,  1979, \mn@doi [\apj] {10.1086/157311}, \href
	{https://ui.adsabs.harvard.edu/abs/1979ApJ...232..520S} {232, 520}
	
	\bibitem[\protect\citeauthoryear{{Timmes} \& {Swesty}}{{Timmes} \&
		{Swesty}}{2000}]{HelmholtzEOS}
	{Timmes} F.~X.,  {Swesty} F.~D.,  2000, \mn@doi [\apjs] {10.1086/313304}, \href
	{https://ui.adsabs.harvard.edu/abs/2000ApJS..126..501T} {126, 501}
	
	\bibitem[\protect\citeauthoryear{{Valtonen} \& {Karttunen}}{{Valtonen} \&
		{Karttunen}}{2006}]{Valtonen+2006}
	{Valtonen} M.,  {Karttunen} H.,  2006, {The Three-Body Problem}
	
	\bibitem[\protect\citeauthoryear{{Vogelsberger} et~al.,}{{Vogelsberger}
		et~al.}{2014}]{Illustris}
	{Vogelsberger} M.,  et~al., 2014, \mn@doi [\mnras] {10.1093/mnras/stu1536},
	\href {https://ui.adsabs.harvard.edu/abs/2014MNRAS.444.1518V} {444, 1518}
	
	\bibitem[\protect\citeauthoryear{{Wang}, {Perna}  \& {Armitage}}{{Wang}
		et~al.}{2021}]{Wang2021TDE}
	{Wang} Y.-H.,  {Perna} R.,   {Armitage} P.~J.,  2021, \mn@doi [\mnras]
	{10.1093/mnras/stab802}, \href
	{https://ui.adsabs.harvard.edu/abs/2021MNRAS.503.6005W} {503, 6005}
	
	\bibitem[\protect\citeauthoryear{{Weinberger}, {Springel}  \&
		{Pakmor}}{{Weinberger} et~al.}{2020}]{Arepo2}
	{Weinberger} R.,  {Springel} V.,   {Pakmor} R.,  2020, \mn@doi [\apjs]
	{10.3847/1538-4365/ab908c}, \href
	{https://ui.adsabs.harvard.edu/abs/2020ApJS..248...32W} {248, 32}
	
	\bibitem[\protect\citeauthoryear{{Yang}, {Bartos}, {Fragione}, {Haiman},
		{Kowalski}, {M{\'a}rka}, {Perna}  \& {Tagawa}}{{Yang}
		et~al.}{2022}]{Yang2022}
	{Yang} Y.,  {Bartos} I.,  {Fragione} G.,  {Haiman} Z.,  {Kowalski} M.,
	{M{\'a}rka} S.,  {Perna} R.,   {Tagawa} H.,  2022, \mn@doi [\apjl]
	{10.3847/2041-8213/ac7c0b}, \href
	{https://ui.adsabs.harvard.edu/abs/2022ApJ...933L..28Y} {933, L28}
	
	\bibitem[\protect\citeauthoryear{{Zwicky}}{{Zwicky}}{1957}]{Zwicky1957}
	{Zwicky} F.,  1957, {Morphological astronomy}
	
	\bibitem[\protect\citeauthoryear{{de Kool}, {van den Heuvel}  \& {Pylyser}}{{de
			Kool} et~al.}{1987}]{deKool1987}
	{de Kool} M.,  {van den Heuvel} E.~P.~J.,   {Pylyser} E.,  1987, \aap, \href
	{https://ui.adsabs.harvard.edu/abs/1987A&A...183...47D} {183, 47}
	
	\makeatother
\end{thebibliography}

\end{document}